\documentclass[superscriptaddress,
twocolumn,secnumarabic,nofootinbib]{revtex4-1}
\usepackage{amsmath}
\usepackage[utf8]{inputenc}
\usepackage[T1]{fontenc}

\pdfoutput=1

\usepackage{bm}
\usepackage{times}
\usepackage{braket}
\usepackage{color,graphicx}
\usepackage[utf8]{inputenc}
\usepackage[T1]{fontenc}
\usepackage[export]{adjustbox}

\usepackage{enumitem}

\usepackage{verbatim}

\usepackage{float}

\usepackage{colortbl}
\usepackage{longtable}
\usepackage{tabularx}
\usepackage{makecell}

\usepackage{multirow}

\usepackage{xcolor}

\usepackage{lipsum}

\usepackage[normalem]{ulem}

\definecolor{LightGray}{gray}{0.96}
\definecolor{Gray}{gray}{0.94}
\definecolor{nicered}{rgb}{0.7,0.1,0.1}
\definecolor{nicegreen}{rgb}{0.1,0.5,0.1}

\usepackage[colorlinks,
            linkcolor=black,
            filecolor=black,
            anchorcolor=black,
            urlcolor=black,
            citecolor=blue,
            bookmarks=false,
            ]{hyperref}

\allowdisplaybreaks

\setlongtables

\begin{document}

\title{Revisiting the $B$-physics anomalies in $R$-parity violating MSSM}

\author{Quan-Yi Hu} \email[]{qyhu@aynu.edu.cn (corresponding author)}
\affiliation{School of Physics and Electrical Engineering, Anyang Normal University, Anyang, Henan 455000, China}
\author{Ya-Dong Yang} \email[]{yangyd@mail.ccnu.edu.cn}
\author{Min-Di Zheng} \email[]{zhengmindi@mails.ccnu.edu.cn}
\affiliation{Institute of Particle Physics and Key Laboratory of Quark and Lepton Physics~(MOE), Central China Normal University, Wuhan, Hubei 430079, China}
  
\begin{abstract}
In recent years, several deviations from the Standard Model predictions in semileptonic decays of $B$-meson might suggest the existence of new physics which would break the lepton-flavour universality. In this work, we have explored the possibility of using muon sneutrinos and right-handed sbottoms to solve these $B$-physics anomalies simultaneously in $R$-parity violating minimal supersymmetric standard model. We find that the photonic penguin induced by exchanging sneutrino can provide sizable lepton flavour universal contribution due to the existence of logarithmic enhancement for the first time. This prompts us to use the two-parameter scenario $(C^{\rm V}_9, \, C^{\rm U}_9)$ to explain $b \to s \ell^+ \ell^-$ anomaly. Finally, the numerical analyses show that the muon sneutrinos and right-handed sbottoms can explain $b \to s \ell^+ \ell^-$ and $R(D^{(\ast)})$ anomalies simultaneously, and satisfy the constraints of other related processes, such as $B \to K^{(\ast)} \nu \bar\nu$ decays, $B_s-\bar B_s$ mixing, $Z$ decays, as well as $D^0 \to \mu^+ \mu^-$, $\tau \to \mu \rho^0$, $B \to \tau \nu$, $D_s \to \tau \nu$, $\tau \to K \nu$, $\tau \to \mu \gamma$, and $\tau \to \mu\mu\mu$ decays. 
\end{abstract}
\pacs{}
\maketitle

\section{Introduction}
\label{sec:intro}

Recently, several flavour anomalies in semileptonic $B$-decays have been reported, which have been attracting great interest. Among them, the observables $R_{K^{(\ast)}} = {\cal B}(B \to K^{(\ast)} \mu^+ \mu^-) / {\cal B}(B \to K^{(\ast)} e^+ e^-)$ in flavour-changing neutral current $b \to s \ell^+ \ell^-$~($\ell = e,\, \mu$) transition and the observables $R(D^{(\ast)}) = {\cal B}(B \to D^{(\ast)} \tau \nu) / {\cal B}(B \to D^{(\ast)} \ell \nu)$ in flavour-changing charged current $b \to c \tau \nu$ transition are particularly striking. The advantage of considering the ratios $R_{K^{(\ast)}}$ and $R(D^{(\ast)})$ instead of the branching fractions themselves is that, apart from the significant reduction of the experimental systematic uncertainties, the Cabibbo-Kobayashi-Maskawa (CKM) matrix elements cancel out and the dependence on the transition form factors become much weaker. These observables can be good probes to test the lepton-flavour universality (LFU) held in the Standard Model (SM).

The latest measurement of $R_K$ by LHCb collaboration gives~\cite{Aaij:2019wad,Aaij:2014ora}
\begin{equation}
\label{eq:RKlhcb}
R_K = 0.846^{+0.060+0.016}_{-0.054-0.014}, \, 1.1 < q^2 < 6        
~{\rm GeV}^2 ,
\end{equation}
but the SM prediction is around 1 with ${\cal O}(1\%)$ uncertainty~\cite{Bordone:2016gaq}, there is $2.5\sigma$ discrepancy. Moreover, the measurement of $R_{K^{\ast}}$ by LHCb at low and high $q^2$ are~\cite{Aaij:2017vbb}
\begin{equation}
\label{eq:RKslhcb}
R_{K^\ast} =
\begin{cases}
0.66^{+0.11}_{-0.07} \pm 0.03 , & 0.045 < q^2 < 1.1 {\rm GeV}^2  \\
0.69^{+0.11}_{-0.07} \pm 0.05 , & 1.1 < q^2 < 6.0 {\rm GeV}^2 
\end{cases} ,
\end{equation}
while the SM predictions are $R_{K^\ast}^{[0.045,\,1.1]} = 0.906 \pm 0.028$ and $R_{K^\ast}^{[1.1,\,6.0]} = 1.00 \pm 0.01$~\cite{Bordone:2016gaq}. The measurements show $2.1\sigma$ discrepancy in the low $q^2$ region and $2.5 \sigma$ discrepancy in the high $q^2$ region, respectively. The Belle collaboration also reported their measurements of $R_{K^{(\ast)}}$~\cite{Abdesselam:2019lab,Abdesselam:2019wac}, which are consistent with the SM predictions within their quite large error bars. In addition to $R_{K^{(\ast)}}$, there are also some other deviations in $b \to s \mu^+ \mu^-$ transition, such as the angular observable $P_5'$~\cite{DescotesGenon:2012zf,Descotes-Genon:2013vna,Hu:2016gpe} of $B \to K^\ast \mu^+ \mu^-$ decay with $2.6 \sigma$ discrepancy~\cite{Aaij:2015oid,Aaij:2013qta,Khachatryan:2015isa,Aaboud:2018krd,Wehle:2016yoi,Abdesselam:2016llu} and the differential branching fraction of $B_s \to \phi \mu^+ \mu^-$ decay with $3.3 \sigma$ discrepancy~\cite{Aaij:2015esa,Aaij:2013aln}.

These deviations indicate the possible existence of new physics (NP) beyond the SM in $b \to s \ell^+ \ell^-$ transition. This NP may break LFU. Many recent model-independent analyses~\cite{Aebischer:2019mlg,Alok:2019ufo,Alguero:2019ptt,Ciuchini:2019usw,Arbey:2019duh,Kowalska:2019ley,Capdevila:2019tsi,Bhattacharya:2019dot} show that some scenarios can explain the $b\to s\ell^+\ell^-$  anomaly well. To express the fit results, we consider the low-energy effective weak Lagrangian governing the $b\to s\ell^+\ell^-$ transition
\begin{equation}
{\cal L}_{\rm eff} =  \frac{4 G_F}{\sqrt{2}} \eta_t \sum_{i}C_i{\cal O}_i + {\rm H.c.} ,
\end{equation} 
where CKM factor $\eta_t \equiv V_{tb} V_{ts}^{\ast}$. We mainly concern the semileptonic operators
\begin{align}
{\cal O}_9 = & \frac{e^2}{16\pi^2}(\bar{s}\gamma_\mu P_{L} b)(\bar{\ell}\gamma^\mu \ell), \\[2mm]   
{\cal O}_{10} = & \frac{e^2}{16\pi^2}(\bar{s}\gamma_\mu P_{L} b)(\bar{\ell}\gamma^\mu \gamma_5 \ell),
\end{align}
where $P_L=(1-\gamma_5)/2$ is the left-handed chirality projector. The Wilson coefficients $C_{9,10} = C_{9,10}^{\rm SM} + C_{9,10}^{\rm NP}$. In this work, we try to explain the anomaly through a two-parameter scenario where the total NP effects are given by~\cite{Alguero:2018nvb}
\begin{align}
C^{\rm NP}_{9,\mu} =  & C^{\rm V}_9 + C^{\rm U}_9, &
C^{\rm NP}_{10,\mu} =  & - C^{\rm V}_9, \\[2mm]
C^{\rm NP}_{9,e} =  & C^{\rm U}_9, &
C^{\rm NP}_{10,e} =  & 0.
\end{align}
The global analyses show that this scenario has the largest pull-value. The best-fit point performed by Ref.~\cite{Alguero:2019ptt} is $(C^{\rm V}_9,\,C^{\rm U}_9) = (-0.30,\,-0.74)$, with the $2\sigma$ range being
\begin{equation}
\label{eq:RPVfit8}
-0.53 < C^{\rm V}_9 < -0.10,\quad
-1.15 < C^{\rm U}_9 < -0.25.
\end{equation}
As we will see in the following discussion, this scenario can be implemented naturally in the $R$-parity violating minimal supersymmetric standard model (MSSM)~\cite{Barbier:2004ez}.

The combined measurements of $R(D^\ast)$ and $R(D)$ are from BaBar~\cite{Lees:2012xj,Lees:2013uzd} and Belle~\cite{Huschle:2015rga,Belle:2019rba}, and Belle~\cite{Hirose:2016wfn,Hirose:2017dxl} and LHCb~\cite{Aaij:2015yra,Aaij:2017uff,Aaij:2017deq} only give the measurements of $R(D^\ast)$. After being averaged by the Heavy Flavor Averaging Group (HFLAV)~\cite{Amhis:2019ckw}, they give the results as follows~\cite{Amhis:2019up}
\begin{align}\label{eq:rd_exp}
R(D)_{\rm avg} = & 0.340 \pm 0.027 \pm 0.013,\\[2mm]
\label{eq:rdstar_exp}
R(D^\ast)_{\rm avg} = & 0.295 \pm 0.011 \pm 0.008,
\end{align}
with a correlation of $-0.38$. Comparing these with the arithmetic average of the SM predictions~\cite{Bigi:2016mdz,Bernlochner:2017jka,Bigi:2017jbd,Jaiswal:2017rve,Amhis:2019up},
\begin{equation}\label{eq:rddstar_theo}
R(D)_{\rm SM}=0.299 \pm 0.003,\quad R(D^\ast)_{\rm SM}=0.258 \pm 0.005,
\end{equation}
one can see that the difference between experiment and theory is at about $3.08\sigma$, implying the existence of LFU violating NP in the charged-current $B$-decays. Global analyses~\cite{Hu:2018veh,Alok:2019uqc,Murgui:2019czp,Shi:2019gxi,Cheung:2020sbq} show that the NP contributing to the left-handed operator $(\bar c \gamma_\mu P_L b)(\bar \tau \gamma^\mu P_L \nu)$ can solve the $R(D^{(\ast)})$ anomaly. Such operator can be generated in $R$-parity violating MSSM by exchanging the right-handed down type squarks at tree level.

There have been attempts to explain the $b \to s \ell^+ \ell^-$ anomaly~\cite{Biswas:2014gga,Das:2017kfo,Earl:2018snx,Darme:2018hqg,Hu:2019ahp} or $R(D^{(\ast)})$ anomaly~\cite{Deshpande:2012rr,Zhu:2016xdg,Altmannshofer:2017poe,Hu:2018lmk,Wang:2019trs} or both of them~\cite{Deshpand:2016cpw,Trifinopoulos:2018rna,Trifinopoulos:2019lyo} by $R$-parity violating interactions in the supersymmetric (SUSY) models. For example, based on the inspiration from the paper by Bauer and Neubert~\cite{Bauer:2015knc}, the authors in Ref.~\cite{Deshpand:2016cpw} investigated the possibility of using right-handed down type squarks to explain the $b \to s \ell^+ \ell^-$ and $R(D^{(\ast)})$ anomalies simultaneously, and found that this was impossible due to the severe constraints from $B\to K^{(\ast)} \nu \bar{\nu}$ decays. Considering that the parameter space obtained by using squarks to explain $b \to s \ell^+ \ell^-$ anomaly is very small~\cite{Deshpand:2016cpw,Das:2017kfo,Earl:2018snx} due to the strict constraints from other related processes, such as $B\to K^{(\ast)} \nu \bar{\nu}$ decays and $B_s-\bar B_s$ mixing, the authors in Ref.~\cite{Hu:2019ahp} used sneutrinos to explain it and found that it is almost unconstrained by other related processes. Based on this knowledge, in this work, we will explore the possibility of using muon sneutrinos $\tilde{\nu}_\mu$ and right-handed sbottoms $\tilde{b}_R$ to explain the $b \to s \ell^+ \ell^-$ and $R(D^{(\ast)})$ anomalies simultaneously within the context of $R$-parity violating MSSM.

Our paper is organized as follows. In Sec.~\ref{sec:b2sll}, we scrutinize all the one-loop contributions of terms $\lambda'_{ijk} L_i Q_j D_k^c$ to $b\to s\ell^+\ell^-$ processes in the framework of $R$-parity violating MSSM, and then give our scenario to explain the $b \to s \ell^+ \ell^-$ anomaly. Discussions of $R(D^{(\ast)})$ anomaly and other related processes are included in Sec.~\ref{sec:constraint}. The numerical analyses and results are shown in Sec.~\ref{sec:result}. Our conclusions are finally made in Sec.~\ref{sec:conclusion}.

\section{$b\to s\ell^+\ell^-$ processes in $R$-parity violating MSSM}
\label{sec:b2sll}

\renewcommand*{\thefootnote}{\arabic{footnote}}

The superpotential terms violating $R$-parity in the MSSM are~\cite{Barbier:2004ez}
\begin{align}
\label{eq:W_RPV}
W_{\rm RPV} = &\, \mu_i L_i H_u + \frac{1}{2} \lambda_{ijk} L_i L_j E_k^c + \lambda'_{ijk} L_i Q_j D_k^c\nonumber \\[2mm]
&+ \frac{1}{2} \lambda''_{ijk} U_i^c D_j^c D_k^c\, ,
\end{align}
where the generation indices are denoted by $i,j,k=1,2,3$ and the colour indices are suppressed. All repeated indices are assumed to be summed over throughout this paper unless otherwise stated (For example, repeated indices in both numerator and denominator are not automatically summed). $H_u$, $L$ and $Q$ are $SU(2)$ doublet chiral superfields while $E^c$, $D^c$ and $U^c$ are $SU(2)$ singlet chiral superfields. 

In this work, we are mainly interested in the terms $\lambda'_{ijk} L_i Q_j D_k^c$ which related to both quarks and leptons. This choice can also alleviate the constraint of sneutrino masses on the collider, because the lower limit of sneutrino masses will be as high as TeV scale~\cite{Aaltonen:2010fv,Abazov:2010km,Aad:2015pfa,Khachatryan:2016ovq} when there are non-zero $\lambda$ and $\lambda'$ at the same time. The corresponding Lagrangian can be obtained by the chiral superfields composing of the fermions and sfermions as follows
\begin{align}
\label{eq:RPVlag}
{\cal L} = &\, \lambda'_{ijk}\big(\tilde{\nu}_{Li} \bar{d}_{Rk} d_{Lj} + \tilde{d}_{Lj} \bar{d}_{Rk} \nu_{Li} + \tilde{d}_{Rk}^\ast \bar{\nu}_{Li}^c d_{Lj}\nonumber \\[2mm] 
&- \tilde{l}_{Li} \bar{d}_{Rk} u_{Lj} - \tilde{u}_{Lj} \bar{d}_{Rk} l_{Li} - \tilde{d}_{Rk}^\ast \bar{l}_{Li}^c u_{Lj}\big) + {\rm H.c.} ,
\end{align}
where the sparticles are denoted by ``$\tilde{\ }$'', and ``c'' indicates charge conjugated fields. Working in the mass eigenstates for the down type quarks and assuming sfermions are in their mass eigenstates, one replaces $u_{Lj}$ by $(V^\dagger u_L)_j$ in Eq.~\eqref{eq:RPVlag}.

These $R$-parity violating interactions can induce $b \to s \ell^+ \ell^-$ processes by exchanging left-handed up squarks $\tilde{u}_{Lj}$ at tree level, but resulting in the operators with right-handed quark current, which are unable to explain the $b \to s \ell^+ \ell^-$ anomaly. This unwanted effect can be eliminated by assuming that the masses of $\tilde{u}_{Lj}$ are very large or/and by assuming that $\lambda'_{ij2} = 0$. Assuming that $\lambda'_{ij2} = 0$ also forbids the exchange of $\tilde{l}_{Li}$ or/and $\tilde{d}_{Lj}$ in one loop level to affect the $b \to s \ell^+ \ell^-$ processes\footnote{In this work, we don't consider contributions only from $R$-parity conserving MSSM, because these contributions can be ignored numerically~\cite{Altmannshofer:2014rta}.}. In the following discussion, we should assume that $\lambda'_{ij1} = \lambda'_{ij2} = 0$.

\begin{figure}[h]
	\centering
	\includegraphics[width=0.42\textwidth]{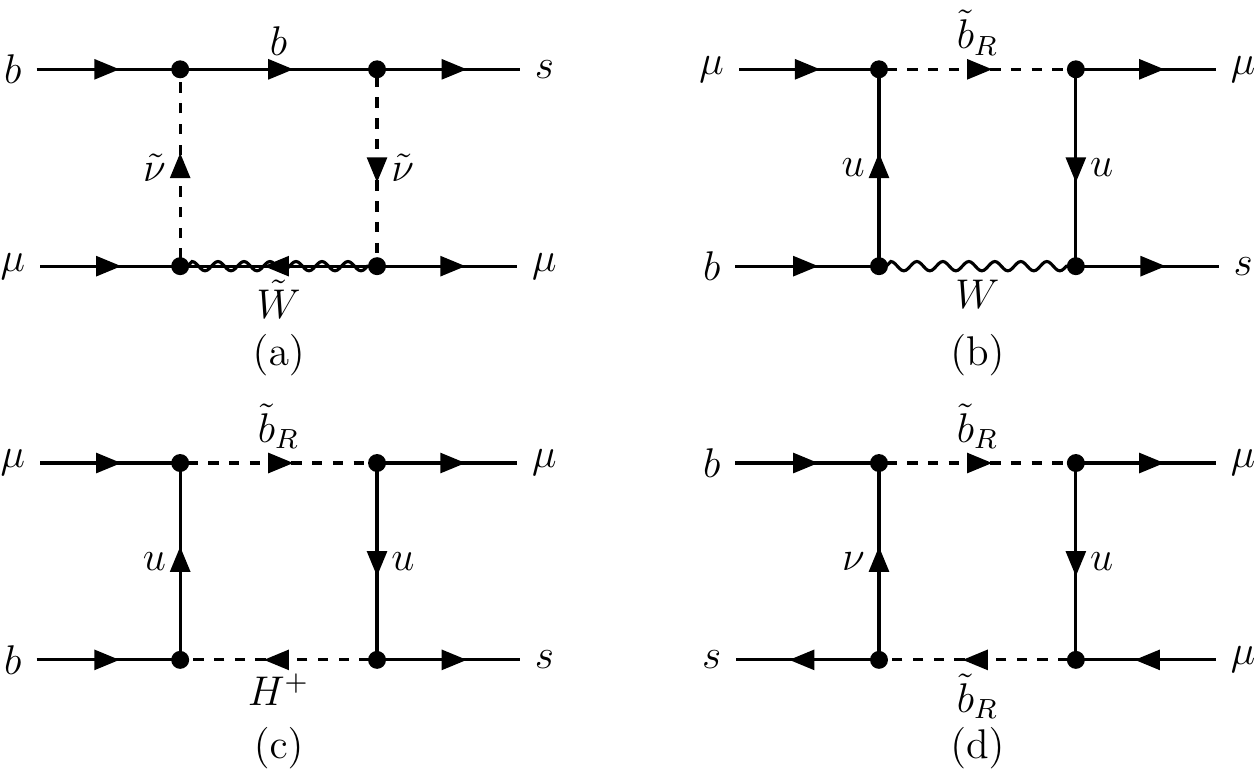}
	\caption{\label{fig:box}Box diagrams for $b \to s \mu^+ \mu^-$ transition in our scenario. Fig.~\ref{fig:box}a shows an example $\tilde{W} - b$ box diagram, Fig.~\ref{fig:box}b shows an example $W - \tilde{b}_R$ box diagram, Fig.~\ref{fig:box}c shows the $H^\pm - \tilde{b}_R$ box diagram, and Fig.~\ref{fig:box}d shows an example $4\lambda'$ box diagram.}
\end{figure}

\begin{figure}[h]
\centering
\includegraphics[width=0.42\textwidth]{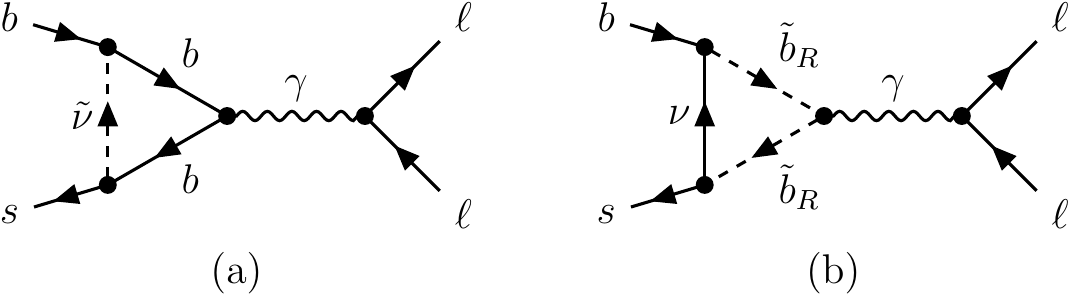}
\caption{\label{fig:penguin}Photonic penguin diagrams studied in our scenario.}
\end{figure}

Next, we will show the contributions of $R$-parity violating MSSM to $b\to s\ell^+\ell^-$ processes. All the Feynman diagrams include four $\tilde{W} - b$ box diagrams (Fig.~\ref{fig:box}a), five $W - \tilde{b}_R$ box diagrams (one of which is Goldstone$- \tilde{b}_R$ box diagram) (Fig.~\ref{fig:box}b), one $H^\pm - \tilde{b}_R$ box diagram (Fig.~\ref{fig:box}c), two $4\lambda'$ box diagrams (Fig.~\ref{fig:box}d) and two $\gamma$-penguin diagrams (Fig.~\ref{fig:penguin}). Most of these results can be found in Refs~\cite{Deshpand:2016cpw,Das:2017kfo,Earl:2018snx,Hu:2019ahp}, however, to our knowledge, the results of the diagram induced by exchanging charged Higgs $H^\pm$ and right-handed sbottom $\tilde{b}_R$ in loop are the first to be given in this paper. The photonic penguin diagrams, which have been neglected in previous work, play an important role in our discussion, as we will explain in more detail later. We do not find sizable $Z$-penguin contributions to $b\to s\ell^+\ell^-$ processes. In this work, the contributions of $\gamma/Z$-penguin diagrams always include their supersymmetric counterparts unless otherwise specified. For convenience, the following Passarino-Veltman functions~\cite{Passarino:1978jh} $D_0$ and $D_2$ are defined as

\begin{align}
&D_0[m_1^2,m_2^2,m_3^2,m_4^2] \notag \\
\equiv& \int \frac{d^4 k}{(2\pi)^4}\frac{1}{(k^2-m_1^2)(k^2-m_2^2)(k^2-m_3^2)(k^2-m_4^2)} \notag \\
=& -\frac{i}{16\pi^2} \biggl[\frac{m_1^2 \log(m_1^2)}{(m_1^2-m_2^2)(m_1^2-m_3^2)(m_1^2-m_4^2)} \notag\\
&+ (m_1 \leftrightarrow m_2) + (m_1 \leftrightarrow m_3) + (m_1 \leftrightarrow m_4)\biggr], \\
&D_2[m_1^2,m_2^2,m_3^2,m_4^2] \notag\\
\equiv& \int \frac{d^4 k}{(2\pi)^4}\frac{k^2}{(k^2-m_1^2)(k^2-m_2^2)(k^2-m_3^2)(k^2-m_4^2)} \notag \\
=& -\frac{i}{16\pi^2} \biggl[\frac{m_1^4 \log(m_1^2)}{(m_1^2-m_2^2)(m_1^2-m_3^2)(m_1^2-m_4^2)} \notag\\
&+ (m_1 \leftrightarrow m_2) + (m_1 \leftrightarrow m_3) + (m_1 \leftrightarrow m_4)\biggr].
\end{align}

The contributions of box diagram are listed below. We eliminate the contributions of all box diagrams to $b \to s e^+ e^-$ processes by assuming $\lambda'_{1j3} = 0$.

\begin{itemize}
	\item The contributions of $\tilde{W} - b$ box diagram to $b \to s \mu^+ \mu^-$ processes are given by
	\begin{align}
	C_9^{{\rm V}(\tilde{W})} = & \frac{- i \pi^2}{\sqrt{2} G_F \sin^2\theta_W \eta_t}\times \notag \\
	&\Bigl( \lambda'_{2i3}\lambda'^*_{223}V_{ib} D_2[m^2_{\tilde{W}},m^2_{\tilde{u}_{Li}},m^2_{\tilde{\nu}_{\mu}},m^2_b] \notag \\
	&-  \lambda'_{2i3}\lambda'^*_{2j3}V_{ib}V^*_{js} D_2[m^2_{\tilde{W}},m^2_{\tilde{u}_{Li}},m^2_{\tilde{u}_{Lj}},m^2_b] \notag \\
	&+  \lambda'_{233}\lambda'^*_{2j3}V^*_{js} D_2[m^2_{\tilde{W}},m^2_{\tilde{u}_{Lj}},m^2_{\tilde{\nu}_{\mu}},m^2_b] \notag \\ 
	& -  \lambda'_{233}\lambda'^*_{223} D_2[m^2_{\tilde{W}},m^2_{\tilde{\nu}_{\mu}},m^2_{\tilde{\nu}_{\mu}},m^2_b] \Bigr),
	\end{align}
	where the winos engage these interactions with left-hand up type squarks and muon sneutrinos. The last term plays an important role in numerical analysis~\cite{Hu:2019ahp}.
	
	\item The contributions of $W - \tilde{b}_R$ box diagram to $b \to s \mu^+ \mu^-$ processes are given by
	\begin{align}
	C_9^{{\rm V}(W)} = &\frac{-i \pi^2}{\sqrt{2}G_F \sin^2\theta_W \eta_t}\times \notag \\
	&\Bigl( \tilde{\lambda}'_{2i3}\lambda'^*_{223}V_{ib} D_2[m^2_{\tilde{b}_{R}},m^2_{u_i},m^2_{W},0] \notag \\
	&-  \tilde{\lambda}'_{2i3}\tilde{\lambda}'^*_{2j3}V_{ib}V^*_{js} D_2[m^2_{\tilde{b}_{R}},m^2_{u_i},m^2_{u_j},m^2_W] \notag \\
	&+  \lambda'_{233}\tilde{\lambda}'^*_{2j3}V^*_{js} D_2[m^2_{\tilde{b}_{R}},m^2_{u_j},m^2_{W},0] \notag \\ 
	&-  \lambda'_{233}\lambda'^*_{223} D_2[m^2_{\tilde{b}_{R}},m^2_W,0,0] \notag \\
	&+ \tilde{\lambda}'_{2i3}\tilde{\lambda}'^*_{2j3}V_{ib}V^*_{js} \frac{m^2_{u_i}m^2_{u_j}}{m_W^2}\notag \\
	&\times D_0[m^2_{\tilde{b}_{R}},m^2_{u_i},m^2_{u_j},m^2_W] \Bigr),
	\end{align}
	where $\tilde{\lambda}'_{ijk} \equiv \lambda'_{ilk}V^{\ast}_{jl}$. The right-hand sbottom $\tilde{b}_{R}$ is the only NP particle here. In the limit $m_{\tilde{b}_R} \gg m_t$, one has $C_9^{{\rm V}(W)}= \frac{m_t^2}{16\pi\alpha m^2_{\tilde{b}_{R}}}|\lambda'_{233}|^2$~\cite{Bauer:2015knc,Earl:2018snx,Das:2017kfo} which is obviously positive.
	
	\item The contributions of $H^\pm - \tilde{b}_R$ box diagram to $b \to s \mu^+ \mu^-$ processes are given by
	\begin{align}
	C_9^{{\rm V}(H^\pm)} =&\frac{-i\pi^2 V_{ib}V_{js}^* \tilde{\lambda}'_{2i3}\tilde{\lambda}'^*_{2j3}}{\sqrt{2} G_F \sin^2\theta_W \tan^2\beta \eta_t}\frac{m_{u_i}^2 m_{u_j}^2}{m_W^2}\notag \\
	&\times D_0[m^2_{H^\pm},m_{u_i}^2,m_{u_j}^2,m^2_{\tilde{b}_{R}}],
	\end{align}
	which should be considered in the following numerical analysis. The $\tan\beta=v_u/v_d$ where $v_u$ and $v_d$ are the vacuum expectation values of two Higgs doublets respectively. 
	
	\item The contributions of $4\lambda'$ box diagram to $b \to s \mu^+ \mu^-$ processes are given by
	\begin{align}
	C_9^{{\rm V}(4\lambda')} =& \frac{-i\pi \lambda'_{i33} \lambda'^*_{i23}}{4\sqrt{2}G_F \alpha \eta_t} \Bigl(|\tilde{\lambda}'_{2j3}|^2  D_2[m^2_{\tilde{b}_{R}},m^2_{\tilde{b}_{R}},m^2_{u_j},0] \notag \\
	&+ |\lambda'_{2j3}|^2  D_2[m^2_{\tilde{u}_{Lj}},m^2_{\tilde{\nu}_i},m^2_b,m^2_b]\Bigr).
	\end{align}
\end{itemize}

The contributions of photonic penguin diagrams are lepton flavour universal which naturally gives us a nonzero $C_9^{\rm U}$
\begin{equation}
\label{eq:phopeng}
C_9^{\rm U} = \frac{\sqrt{2}\lambda'_{i33}\lambda'^*_{i23}}{36 G_F \eta_t}\biggl[ \frac{1}{6 m_{\tilde{b}_{R}}^2} -\biggl(\frac{4}{3} + \log\frac{m_b^2}{m_{\tilde{\nu}_i}^2}  \biggr)\frac{1}{m_{\tilde{\nu}_i}^2}\biggr].
\end{equation}
As stated in Ref.~\cite{Hu:2019ahp}, this result is consistent with that in Ref.~\cite{deGouvea:2000cf}, but it has a {\it negative} sign different from that in Ref.~\cite{Earl:2018snx}. The first term in Eq.~\eqref{eq:phopeng} comes from the contribution of Fig.~\ref{fig:penguin}b, like the photonic penguin induced by scalar leptoquark. We find this term gives a negligible contribution, which is in agreement with Refs.~\cite{Bauer:2015knc,Das:2016vkr}. However the second term in Eq.~\eqref{eq:phopeng} has a significant contribution because of the logarithmic enhancement, which has never been addressed before. These photonic penguins also contribute new electromagnetic dipole operator ${\cal O}_7=\frac{m_b}{e}(\bar s \sigma^{\alpha \beta} P_R b)F_{\alpha \beta}$, which is strictly constrained by $B \to X_s \gamma$ decay~\cite{Hu:2016gpe}. Fortunately, we find that the corresponding contribution can be ignored numerically because there such logarithmic enhancement is absent~\cite{Earl:2018snx,Hu:2019ahp,deGouvea:2000cf}.

We will discuss the possibility of using muon sneutrinos $\tilde{\nu}_\mu$ and right-handed sbottoms $\tilde{b}_R$ to explain $b \to s \ell^+ \ell^-$ anomaly, for which we set the mass of tau sneutrinos $\tilde{\nu}_\tau$ and three left-handed up type squarks $\tilde{u}_{Lj}$ sufficiently large that the contributions of the loop diagrams containing them are ignored\footnote{In our numerical analysis, we find that the contribution of the loop diagrams containing $\tilde{\nu}_\tau$ is numerically negligible when the mass of $\tilde{\nu}_\tau$ is a few TeV or larger. The same conclusion is true for $\tilde{u}_{L}$ where the mass of $\tilde{u}_{L}$ is a few 10TeV or larger. Here, we consider that $m_{\tilde{\nu}_\mu} < m_{\tilde{\nu}_\tau}$, which can be achieved, for example, by setting the hierarchy of neutrino Yukawas $Y_{\nu_2}<Y_{\nu_3}$ in the $\mu\nu$SSM~\cite{Kpatcha:2019pve}.}. The contribution from $H^\pm - \tilde{b}_R$ box diagram is usually positive, and we find that it is numerically negligible when $\tan\beta>2$. Thus, the contributions to only muon channel are 
\begin{align}
C_9^{\rm V} =& -\frac{\sqrt{2} \lambda'_{233}\lambda'^\ast_{223}f(x_{\tilde{\nu}_\mu})}{32 G_F \sin^2\theta_W \eta_t m^2_{\tilde{\nu}_\mu}} + \frac{|\lambda'_{233}|^2 x_{\tilde{b}_R}}{16\pi\alpha} 
\\ 
&-\frac{\lambda'_{i33}\lambda^{\prime \ast}_{i23} \left[|\tilde{\lambda}'_{213}|^2 + |\tilde{\lambda}'_{223}|^2 - |\tilde{\lambda}'_{233}|^2 f\left(1/x_{\tilde{b}_R}\right)\right]  }{64\sqrt{2}\pi G_F \alpha \eta_t m^2_{\tilde{b}_R}},\notag 
\end{align}
where $x_{\tilde{\nu}_\mu} \equiv m^2_{\tilde{\nu}_\mu}/m^2_{\tilde{W}}$, $x_{\tilde{b}_R} \equiv m^2_t/m^2_{\tilde{b}_R}$, and the loop function $f(x) \equiv \frac{x (1-x+\log{x}) }{(1-x)^2}$.

\section{$R(D^{(\ast)})$ anomaly and other constraints}
\label{sec:constraint}

In this section, we discuss the interpretation of $R(D^{(\ast)})$ anomaly and consider the constraints imposed by other related processes from $B,\,D,\,K,\,\tau$, and $Z$ decays.

\subsection{$R(D^{(\ast)})$ anomaly}

In $R$-parity violating MSSM, the charged current processes $d_j \to u_n l_l\nu_i$ are induced by exchanging $\tilde{b}_R$ at tree level. The effective Lagrangian of these processes are given by
\begin{equation}
{\cal L}_{\rm eff} = -\frac{4G_F}{\sqrt{2}} V_{nj}
(\delta_{li}+C_{njli}) \bar{u}_n \gamma_{\mu} P_L d_j \bar{l}_l \gamma^{\mu} P_L \nu_i + {\rm H.c.}, 
\end{equation} 
where the Wilson coefficient $C_{njli}$ is 
\begin{equation}
C_{njli} = \frac{\lambda'_{ij3} \tilde{\lambda}^{\prime\ast}_{ln3}}{4 \sqrt{2} G_F V_{nj} m_{\tilde{b}_{R}}^2} .
\end{equation}
Because taking $\lambda'_{1j3}=0$ to eliminate the contributions of box diagrams to $b \to s e^+ e^-$ processes\footnote{In fact, by combining the assumptions $\lambda'_{1j3}=0$ and $\lambda'_{ij1}=\lambda'_{ij2}=0$, we can get $\lambda'_{1jk}=0$, which implies that the contribution of box diagrams of NP to the first generation leptons and sleptons is zero, because we only consider the terms $\lambda'_{ijk} L_i Q_j D_k^c$.}, we have $C_{nj1i} = C_{njl1} = 0$. It is useful to define the ratio
\begin{equation}
R_{njl}\equiv \frac{{\cal B}(d_j\to  u_n l_l \nu)}{{\cal B}(d_j\to u_n l_l \nu)_{\rm SM}} =\sum_{i=1}^3 \left|\delta_{li}+C_{njli}\right|^2,
\end{equation}
and we have 
\begin{equation}
\frac{R(D)}{R(D)_{\rm SM}} =\frac{R(D^\ast)}{R(D^\ast)_{\rm SM}} = \frac{2R_{233}}{R_{232} + 1}.
\end{equation}
To obtain the allowed parameter region, we use the following best fit value in the $R$-parity violating scenario
\begin{equation}
\frac{R(D)}{R(D)_{\rm SM}} =\frac{R(D^\ast)}{R(D^\ast)_{\rm SM}} = 1.14 \pm 0.04.
\end{equation}

\subsection{Constraints from the tree-level processes}

In the scenario we set up, some other processes receive tree level $R$-parity violating contributions. Here we mainly discuss the constraints from neutral current processes $B \to K^{(\ast)} \nu \bar\nu$, $B \to \pi \nu \bar\nu$, $K \to \pi \nu \bar\nu$, $D^0 \to \mu^+ \mu^-$ and $\tau \to \mu \rho^0$, as well as charged current processes $B \to \tau \nu$, $D_s \to \tau \nu$ and $\tau \to K \nu$. These decays relate to
\begin{align}
\frac{\lambda'_{ij3}\lambda^{\prime \ast}_{lm3}}{2 m^2_{\tilde{b}_R}} \bar{d}_m \gamma^\mu P_L d_j \bar{\nu}_l\gamma_\mu P_L \nu_i,
\\
\frac{\tilde{\lambda}'_{ij3}\tilde{\lambda}^{\prime \ast}_{lm3}}{2 m^2_{\tilde{b}_R}} \bar{u}_m \gamma^\mu P_L  u_j \bar{l}_l\gamma_\mu P_L l_i,
\\
-\frac{\lambda'_{ij3}\tilde{\lambda}^{\prime \ast}_{lm3}}{2 m^2_{\tilde{b}_R}} \bar{u}_m \gamma^\mu P_L d_j \bar{l}_l \gamma_\mu P_L \nu_i.
\end{align}

The effective Lagrangian for  $B \to K^{(\ast)} \nu \bar\nu$, $B \to \pi \nu \bar\nu$ and $K \to \pi \nu \bar\nu$ decays are defined by
\begin{equation}
\label{eq:ddnunuleff}
{\cal L}_{\rm eff} = (C_{mj}^{\rm SM} \delta_{li} + C_{mj}^{\nu_l \bar{\nu}_i }) ({\bar d}_m \gamma^\mu P_L d_j)({\bar \nu}_l \gamma_\mu P_L \nu_i) + {\rm H.c.},
\end{equation}
where~\cite{Buras:2014fpa}
\begin{equation}
C_{mj}^{\rm SM} = - \frac{\sqrt{2}G_F \alpha X(x_t)}{\pi \sin^2\theta_W} V_{tj} V^\ast_{tm},
\end{equation}
is the SM one. The loop function $X(x_t)\equiv \frac{x_t(x_t+2)}{8(x_t-1)}+\frac{3 x_t (x_t-2)}{8(x_t-1)^2}\log(x_t)$ with $x_t \equiv m^2_t/m^2_W$. The $R$-parity violating contributions are given by
\begin{equation}
C_{mj}^{\nu_l \bar{\nu}_i } = \frac{\lambda'_{ij3}\lambda^{\prime \ast}_{lm3}}{2 m^2_{\tilde{b}_R}}.
\end{equation}
It is useful to define the ratio 
\begin{align}
\label{eq:Rbsnnu}
R^{\nu \bar \nu}_{mj } \equiv& \frac{{\cal B}(d_j\to d_m \nu \bar{\nu})}{{\cal B}(d_j\to d_m \nu \bar{\nu})_{\rm SM}} \notag \\
=& \frac{\sum\limits_{i=1}^3 \left| C_{mj}^{\rm SM} + C_{mj}^{\nu_i \bar{\nu}_i }\right|^2 + \sum\limits_{i \neq l}^3 \left| C_{mj}^{\nu_l \bar{\nu}_i } \right|^2}{3 \left| C_{mj}^{\rm SM} \right|^2}.
\end{align}
The upper limit of $B \to K^{(\ast)} \nu \bar\nu$ decay corresponds to $R^{\nu \bar \nu}_{23} < 2.7$~\cite{Grygier:2017tzo,Buras:2014fpa,Lees:2013kla} at 90\% confidence level (CL), and the upper limit of $B \to \pi \nu \bar\nu$ decay is related to $R^{\nu \bar \nu}_{13} < 830.5$~\cite{Lutz:2013ftz,Du:2015tda} at 90\% CL. By combining the SM prediction ${\cal B}(K^+ \to \pi^+\nu\bar\nu)_{\rm SM} = (9.24 \pm 0.83) \times 10^{-11}$~\cite{Aebischer:2018iyb} with experimental measurement ${\cal B}(K^+ \to \pi^+\nu\bar\nu)_{\rm exp} = (1.7 \pm 1.1) \times 10^{-10}$~\cite{Tanabashi:2018oca}, we obtain a stringent constraint from $K \to \pi \nu \bar\nu$ decay that makes 
\begin{equation}
|\lambda'_{i23}\lambda^{\prime \ast}_{l13}| < 7.4\times 10^{-4}(m_{\tilde{b}_R}/1{\rm TeV})^2.
\end{equation}
Therefore, we will assume $\lambda'_{i1k} =0$ to satisfy this constraint. At the same time, under this assumption, $B \to \pi \nu \bar\nu$ decay is unaffected by the NP.

The branching fraction for $D^0 \to \mu^+ \mu^-$ decay is given by~\cite{Deshpand:2016cpw} 
\begin{equation}
{\cal B}(D^0 \to \mu^+ \mu^-) = \frac{\tau_D f^2_D m_D m^2_\mu}{32\pi} \left|\frac{\tilde{\lambda}'_{223}\tilde{\lambda}^{\prime \ast}_{213}}{2 m^2_{\tilde{b}_R}}\right|^2  \sqrt{1-\frac{4 m^2_\mu}{m^2_D}},
\end{equation}
where decay constant of $D^0$ is $f_D = 209.0 \pm 2.4$ MeV~\cite{Aoki:2019cca}. The mean life $\tau_D= 410.1\pm 1.5$ fs~\cite{Tanabashi:2018oca} and the upper limit of branching fraction of $D^0 \to \mu^+ \mu^-$ decay is $6.2 \times 10^{-9}$ at 90\% CL~\cite{Tanabashi:2018oca}. The corresponding constraint is $|\lambda'_{223}|^2 < 0.31(m_{\tilde{b}_R}/1{\rm TeV})^2$.

The branching fraction for $\tau \to \mu \rho^0$ decay is given by~\cite{Kim:1997rr}
\begin{align}
{\cal B}(\tau \to \mu \rho^0) = &
\frac{\tau_\tau f^2_\rho m^3_\tau}{128\pi}\left|\frac{\tilde{\lambda}'_{313}\tilde{\lambda}^{\prime \ast}_{213}}{2 m^2_{\tilde{b}_R}}\right|^2 \left(1-\frac{m^2_\rho}{m^2_\tau}\right)\notag \\
&\times \left(1+ \frac{m^2_\rho}{m^2_\tau} - 2\frac{m^4_\rho}{m^4_\tau}\right),
\end{align}
where $\tau_\tau = 290.3 \pm 0.5$ fs and the decay constant $f_\rho = 153$ MeV~\cite{Earl:2018snx}. The current experimental upper limit on the branching fraction for this process is ${\cal B}(\tau \to \mu \rho^0) < 1.2\times 10^{-8}$ at 90\% CL~\cite{Tanabashi:2018oca}. The corresponding constraint is $|\lambda'_{323}\lambda'^*_{223}| < 0.38(m_{\tilde{b}_R}/1{\rm TeV})^2$.

The formulas for charged current processes are given, respectively, by
\begin{align}
\frac{{\cal B}(B \to \tau \nu)}{{\cal B}(B \to \tau \nu)_{\rm SM}} = R_{133}, \\
\frac{{\cal B}(D_s \to \tau \nu)}{{\cal B}(D_s \to \tau \nu)_{\rm SM}} = R_{223},\\
\frac{{\cal B}(\tau \to K \nu)}{{\cal B}(\tau \to K \nu)_{\rm SM}} = R_{123}.
\end{align}
The corresponding experimental and theoretical values are listed, respectively, as follows: ${\cal B}(B \to \tau \nu)_{\rm exp} = (1.09\pm0.24) \times 10^{-4}$~\cite{Tanabashi:2018oca}, ${\cal B}(B \to \tau \nu)_{\rm SM} = (9.47\pm1.82) \times 10^{-5}$~\cite{Nandi:2016wlp}; ${\cal B}(D_s \to \tau \nu)_{\rm exp} = (5.48 \pm 0.23)\%$~\cite{Tanabashi:2018oca}, ${\cal B}(D_s \to \tau \nu)_{\rm SM} = (5.40 \pm 0.30)\%$; ${\cal B}(\tau \to K \nu)_{\rm exp} = (6.96 \pm 0.10)\times 10^{-3}$~\cite{Tanabashi:2018oca}, ${\cal B}(\tau \to K \nu)_{\rm SM} = (7.15 \pm 0.026)\times 10^{-3}$~\cite{Hu:2018lmk}.

\subsection{Constraints from the loop-level processes}

First of all, the most important one-loop constraint comes from $B_s-\bar{B}_s$ mixing, which is governed by
\begin{equation}
\label{eq:Bmixing}
{\cal L}_{\rm eff} = (C_{B_s}^{\rm SM} + C_{B_s}^{\rm NP})(\bar{s}\gamma_{\mu}P_L b)(\bar{s}\gamma^{\mu} P_L b) + {\rm H.c.},
\end{equation}
where the SM and NP Wilson coefficients are given respectively by

\begin{align}
C_{B_s}^{\rm SM} =& -\frac{1}{4 \pi^2} G_F^2 m_W^2 \eta_t^2 S(x_t), \\
C_{B_s}^{\rm NP} =& -\frac{1}{128 \pi^2}\biggl[\frac{(\lambda^{\prime}_{i33} \lambda^{\prime\ast}_{i23})^2 }{m_{\tilde{b}_R}^2}  
+\frac{ (\lambda^{\prime}_{233} \lambda^{\prime\ast}_{223})^2 }{m_{\tilde{\nu}_{\mu}}^2}\biggr],
\end{align}
where loop function $S(x_t)=\frac{x_t(4-11x_t+x_t^2)}{4(x_t-1)^2}+\frac{3x_t^3\log(x_t)}{2(x_t-1)^3}$. At $2\sigma$ level, the UT$fit$ collaboration~\cite{Bona:2007vi} gives the bound $0.93 < |1+ C_{B_s}^{\rm NP}/C_{B_s}^{\rm SM}| < 1.29$.

Next, we investigate a series of $Z$ decaying to two charged leptons with the same flavour like $Z\to \mu\mu(\tau\tau)$ and the different one like $Z\rightarrow \mu\tau$. The amplitude of these diagrams is 
$
i \mathcal M =i \frac{g}{32\pi^2 \cos \theta_W} B_{ij} \epsilon^\alpha \bar{u}_{\ell_i}\gamma_\alpha P_L v_{\ell_j}
$~\cite{Earl:2018snx}, 
where $B_{ij} = B^1_{ij} + B^2_{ij}$ and~\cite{Earl:2018snx,Arnan:2019olv}
\begin{align}
B^1_{ij} =& \ \sum_{l=1}^2 \tilde{\lambda}'_{jl3} \tilde{\lambda}'^{\ast}_{il3} \frac{m_Z^2}{m^2_{\tilde{b}_R}} \biggl[\biggl(1 - \frac{4}{3} \sin^2\theta_W \biggr)\notag\\
&\times \biggl(\log\frac{m_Z^2}{m^2_{\tilde{b}_R}} - i \pi - \frac{1}{3} \biggr) + \frac{\sin^2\theta_W}{9} \biggr], \\
B^2_{ij} =& \ 3 \tilde{\lambda}'_{j33} \tilde{\lambda}'^{\ast}_{i33} \biggl\{-x_{\tilde{b}_R}(1 + \log x_{\tilde{b}_R}) \notag \\
&+ \frac{m_Z^2}{18 m^2_{\tilde{b}_R}}\biggl[(11 - 10 \sin^2\theta_W) + (6 - 8 \sin^2\theta_W)\log x_{\tilde{b}_R}\notag \\ 
&+ \frac{1}{10}(-9 + 16 \sin^2\theta_W)\frac{m_Z^2}{m_t^2} \biggr] \biggl\},
\end{align}
here $B^1_{ij}$ is the contribution from the diagram induced by exchanging $\tilde{b}_R-u-u$ or $\tilde{b}_R-c-c$ in triangular loop and $B^2_{ij}$ is the contribution from the diagram induced by exchanging $\tilde{b}_R-t-t$ in triangular loop. As shown in Ref.~\cite{Earl:2018snx}, for $Z\to \mu\mu(\tau\tau)$, demanding the interference term in the partial width between the SM tree-level contribution and the NP one-loop level ones is less than twice the experimental uncertainty on the partial width~\cite{Tanabashi:2018oca}, there are the bounds $|\Re(B_{22})| < 0.32$ and $|\Re(B_{33})| < 0.39$~\cite{Earl:2018snx}. And the experimental upper limit ${\mathcal B}(Z\rightarrow \mu\tau)<1.2\times 10^{-5}$~\cite{Tanabashi:2018oca} makes the bound $\sqrt{|B_{23}|^2 + |B_{32}|^2} < 2.1$~\cite{Earl:2018snx}.

Finally, we discuss the lepton-flavour violating decay of $\tau$ lepton, including $\tau \to \mu \gamma$ and $\tau \to \mu\mu\mu$. In the limit $m^2_\mu/m^2_\tau \to 0$, the branching fraction for $\tau \to \mu \gamma$ is given by~\cite{Kuno:1999jp,Farzan:2010nh,deGouvea:2000cf}
\begin{equation}
{\cal B}(\tau \to \mu \gamma) = \frac{\tau_\tau \alpha m^5_\tau}{4}(|A^L_2|^2 + |A^R_2|^2),
\end{equation}
where the effective couplings $A^{L,R}_2$ come from on shell photon penguin diagrams~\cite{deGouvea:2000cf},
\begin{equation}
A^L_2 = -\frac{\lambda'_{2j3}\lambda'^*_{3j3}}{64 \pi^2 m^2_{\tilde{b}_R}},
\quad
A^R_2 = 0.
\end{equation}
The current experimental upper limit is ${\cal B}(\tau \to \mu \gamma) < 4.4 \times 10^{-8}$ at 90\% CL~\cite{Tanabashi:2018oca}.

In general, the effective Lagrangian leading to $\tau \to \mu\mu\mu$ decay is given by~\cite{Kuno:1999jp,Farzan:2010nh}
\begin{align}
{\cal L}_{\rm eff}=&
-\frac{B_1}{2}(\bar{\tau} \gamma^\nu P_L \mu)(
\bar{\mu}  \gamma_\nu P_R \mu)-\frac{B_2}{2}(\bar{\tau} \gamma^\nu P_R \mu)(\bar{\mu} \gamma_\nu P_L \mu) \notag\\ 
&+ C_1 (\bar{\tau} P_R \mu)(\bar{\mu} P_R \mu)+ C_2 (\bar{\tau} P_L \mu)(\bar{\mu} P_L \mu) \notag\\ 
&+ G_1 (\bar{\tau} \gamma^\nu P_R \mu)(\bar{\mu} \gamma_\nu
P_R \mu)+ G_2 (\bar{\tau} \gamma^\nu P_L \mu)(\bar{\mu} \gamma_\nu P_L \mu) \notag\\  
&-A_R (\bar{\tau} [\gamma_\mu,\gamma_\nu]\frac{q^\nu}{q^2}    P_R \mu) (\bar{\mu} \gamma^\mu \mu)\notag \\
&-A_L (\bar{\tau} [\gamma_\mu,\gamma_\nu] \frac{q^\nu}{q^2} P_L \mu)(\bar{\mu}\gamma^\mu \mu)+ {\rm H.c.}.
\end{align}
This Lagrangian leads to 
~\cite{Kuno:1999jp,Farzan:2010nh}
\begin{align}\label{Brtaumumumu}
{\cal B}&(\tau \to 3\mu)=
\frac{\tau_\tau m^5_\tau}{6144 \pi^3}
\biggl[ |B_1|^2+|B_2|^2+8(|G_1|^2+|G_2|^2) 
\notag\\ 
&+\frac{|C_1|^2+|C_2|^2}{2}+32\biggl(4\log \frac{m_\tau^2}{
	m_\mu^2} - 11\biggr)\frac{|A_R|^2+|A_L|^2 }{ m_\tau^2}
\notag\\ 
&-64 \frac{\Re(A_L G_2^\ast+A_R G_1^\ast)}{
	m_\tau}+32\frac{\Re(A_L B_1^\ast+A_R B_2^\ast)}{m_\tau} \biggr].
\end{align} 
In our scenario, there are three different types of contributions, the photonic and $Z$ penguins as well as box diagrams with four $\lambda'$ couplings, that can contribute to $\tau \to \mu\mu\mu$ decay. The nonzero Wilson coefficients are~\cite{Earl:2018snx,deGouvea:2000cf}
\begin{align}
B_1=&-2 \bigl(4\pi\alpha A^L_1 + \sin^2\theta_W B' \bigr), \\ 
G_2=& 4\pi\alpha A^L_1 + \biggl(-\frac{1}{2} + \sin^2\theta_W\biggr) B' + C_\tau, \\ 
A_L=& 2\pi\alpha m_\tau A^L_2,
\end{align}
where
\begin{align}
B'=&-\frac{3\alpha \tilde{\lambda}'_{233} \tilde{\lambda}'^{\ast}_{333} x_{\tilde{b}_R}(1 + \log x_{\tilde{b}_R})}{8\pi \cos^2\theta_W \sin^2\theta_W m^2_Z}, \\
C_\tau=& \frac{i}{4} \tilde{\lambda}'_{2i3} \tilde{\lambda}^{'\ast}_{2i3} \tilde{\lambda}'_{2j3} \tilde{\lambda}^{'\ast}_{3j3}  D_2[m^2_{\tilde{b}_R},m^2_{\tilde{b}_R},m^2_{u_i},m^2_{u_j}], 
\end{align}
and the off-shell effective coupling $A^L_1$ is~\cite{deGouvea:2000cf} 
\begin{equation}
A^L_1 = \frac{\lambda'_{2j3}\lambda'^*_{3j3}}{16 \pi^2 m^2_{\tilde{b}_R}} \biggl[ \frac{1}{18} -\frac{2}{3}\biggl(\frac{4}{3} + \log\frac{m^2_{u_j}}{m^2_{\tilde{b}_R}}\biggr)\biggr].
\end{equation}
The current experimental upper limit on the branching fraction for this decay is ${\cal B}(\tau \to \mu\mu\mu) < 2.1\times 10^{-8}$ at 90\% CL~\cite{Tanabashi:2018oca}.

\begin{figure*}[t]
	\centering
	\includegraphics[width=0.32\textwidth]{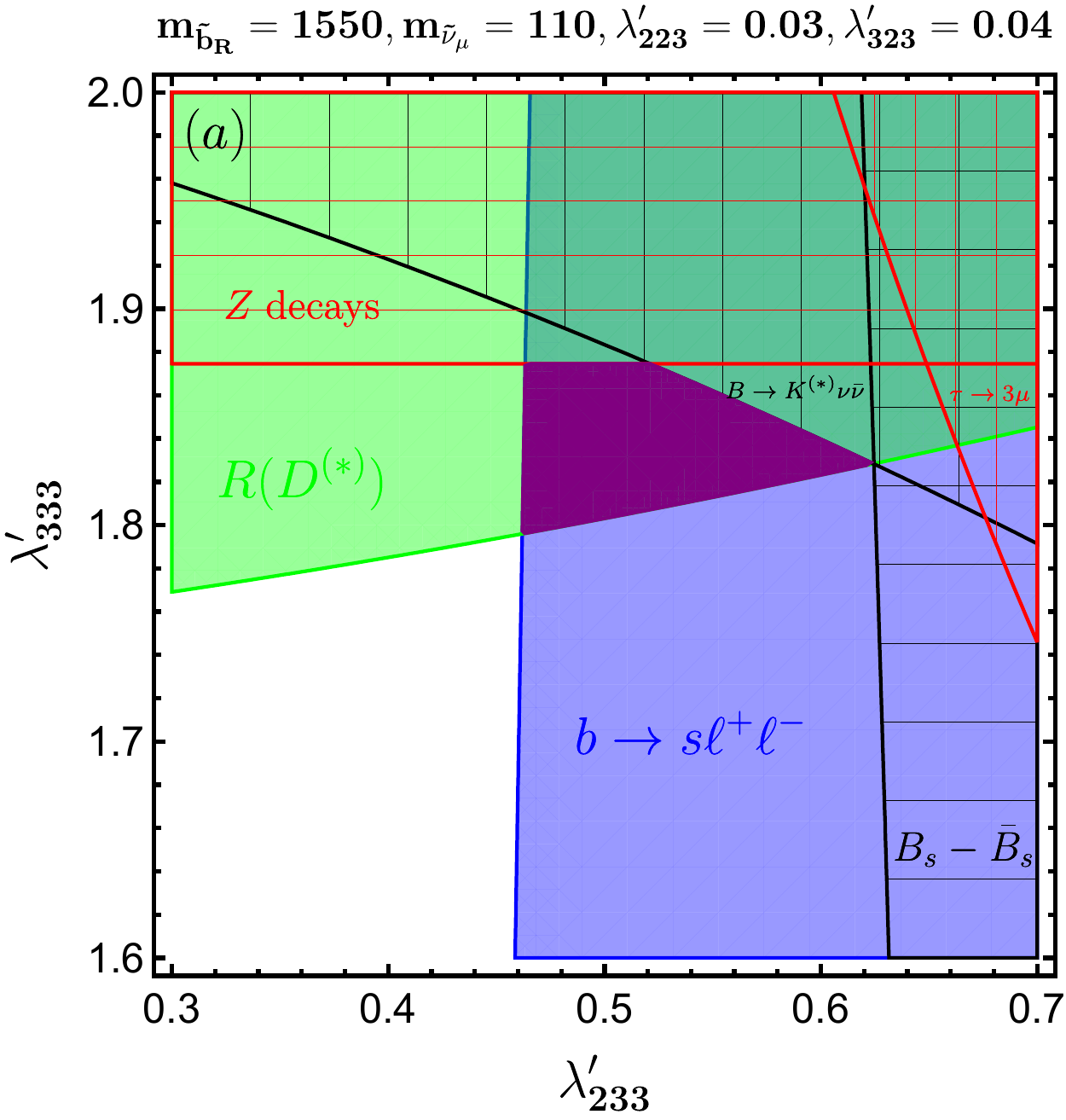}\quad
	\includegraphics[width=0.32\textwidth]{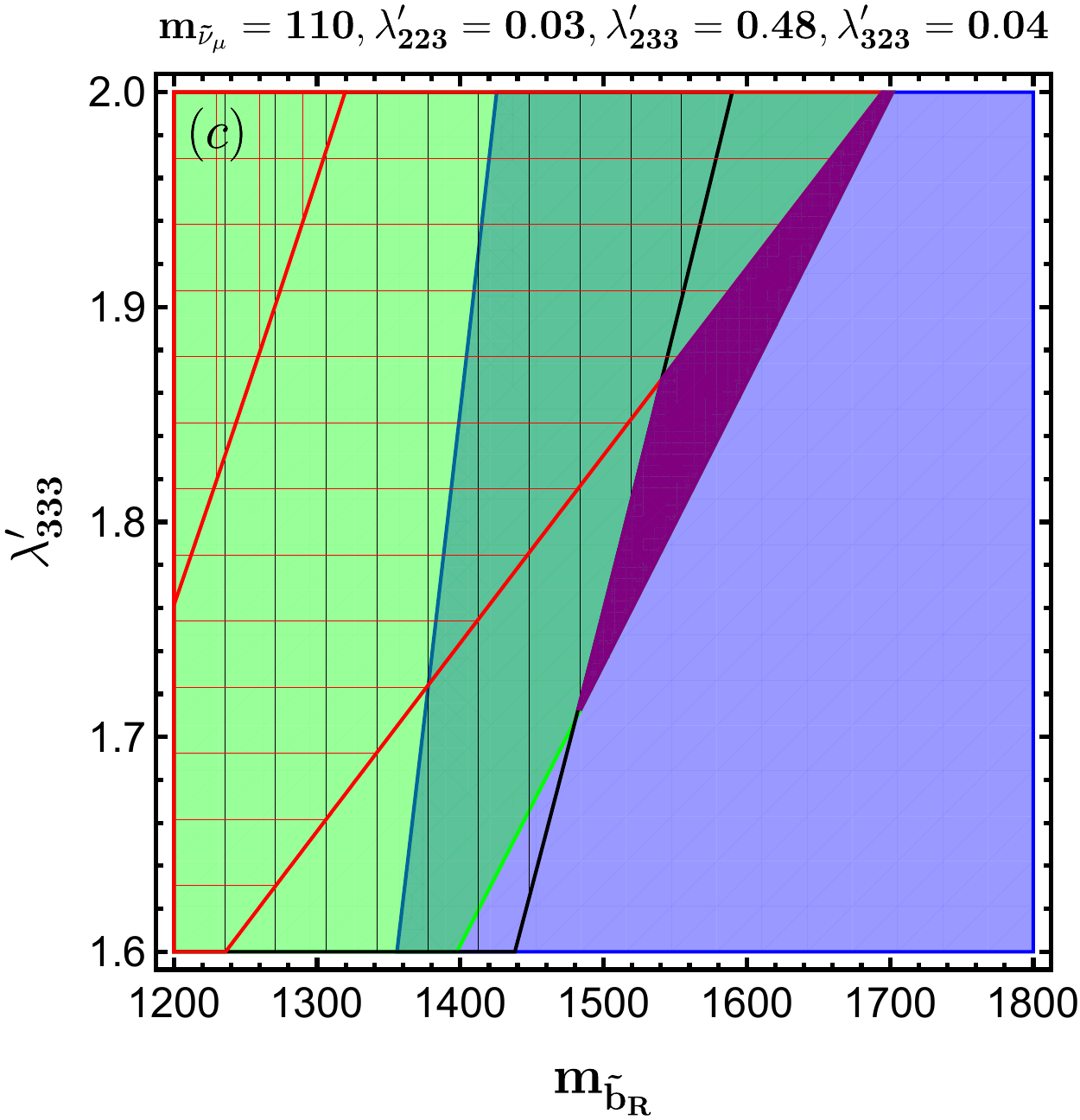}\quad
	\includegraphics[width=0.32\textwidth]{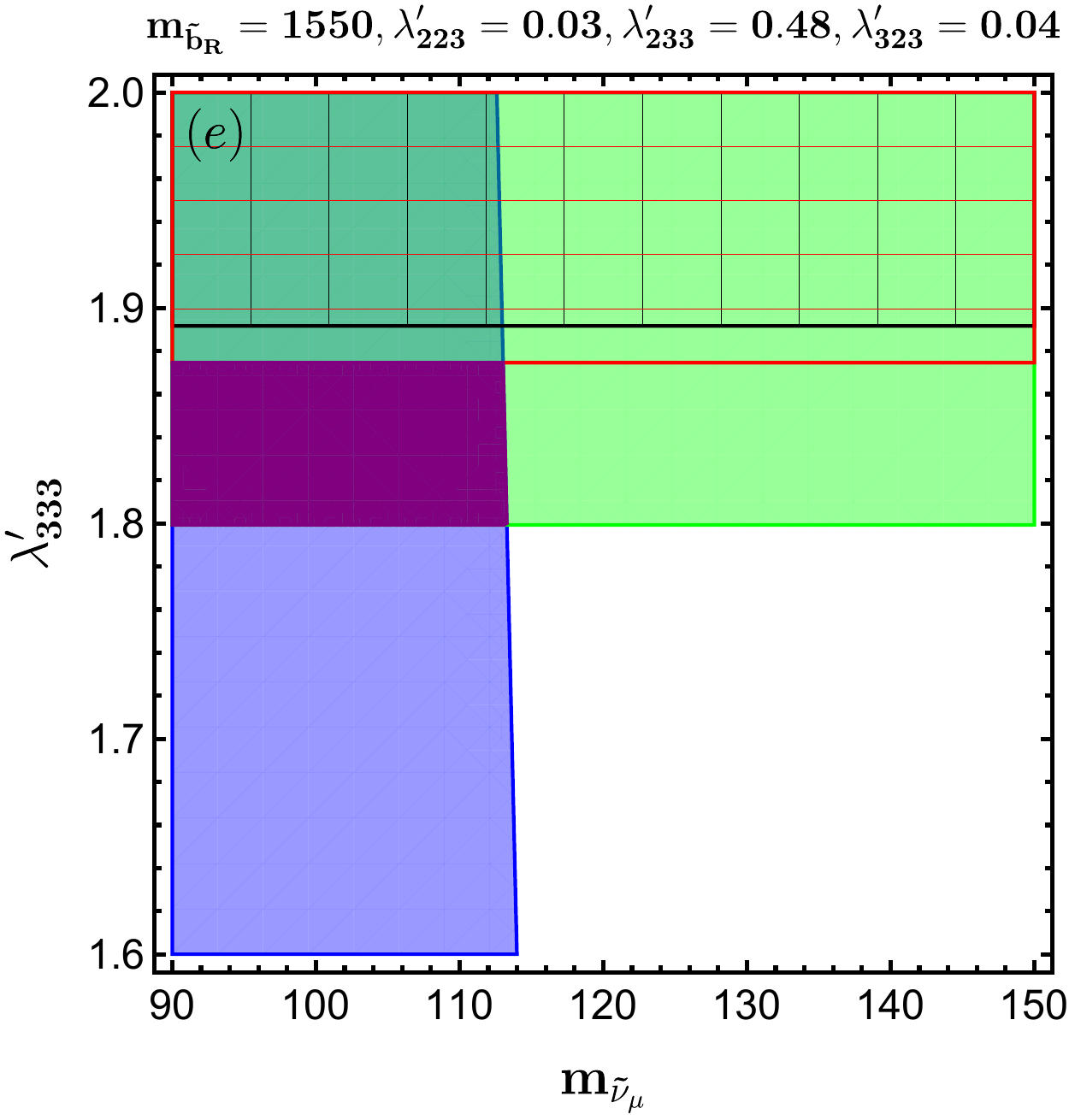}\\[2mm]
	\includegraphics[width=0.32\textwidth]{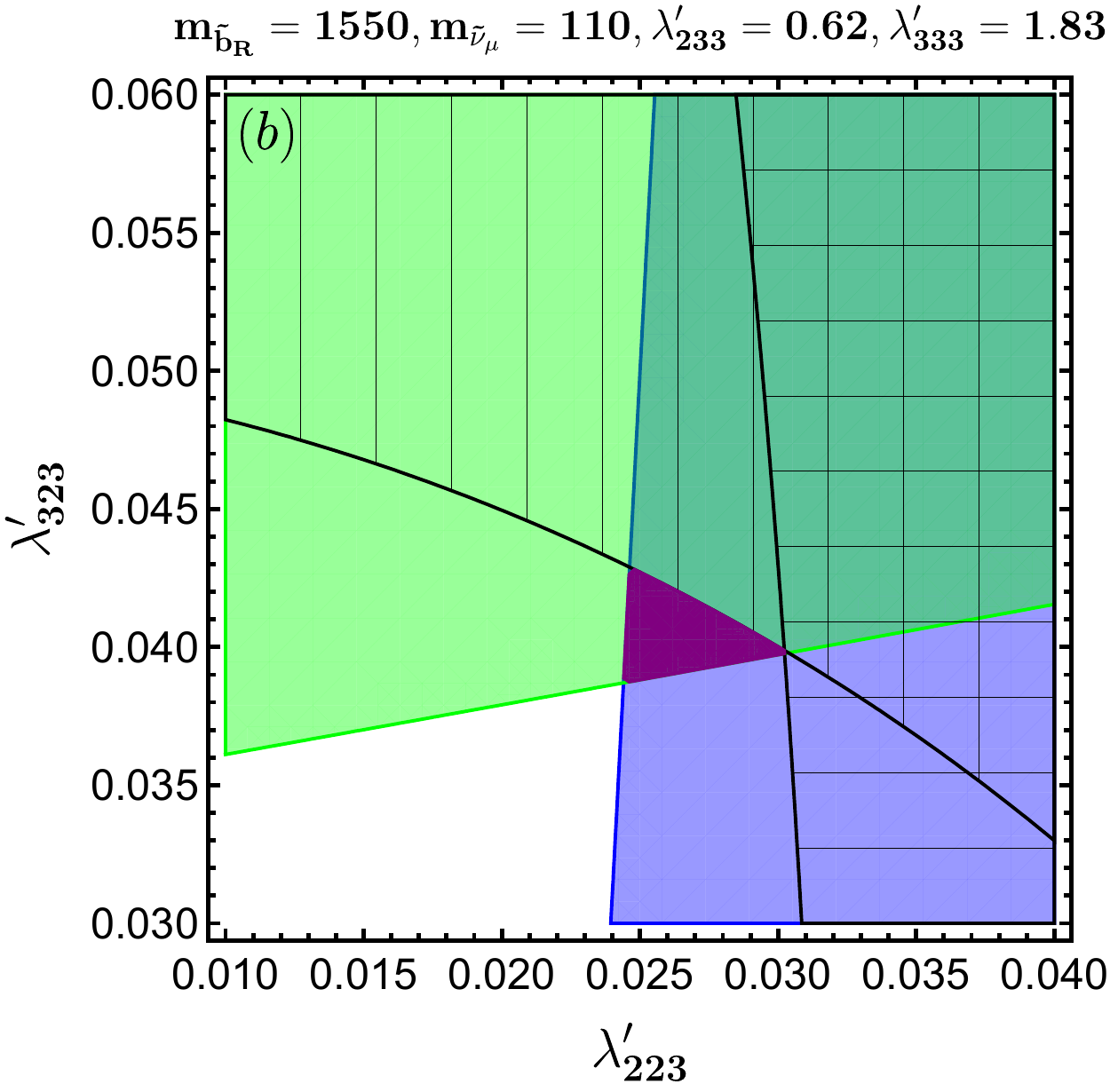}\quad
	\includegraphics[width=0.32\textwidth]{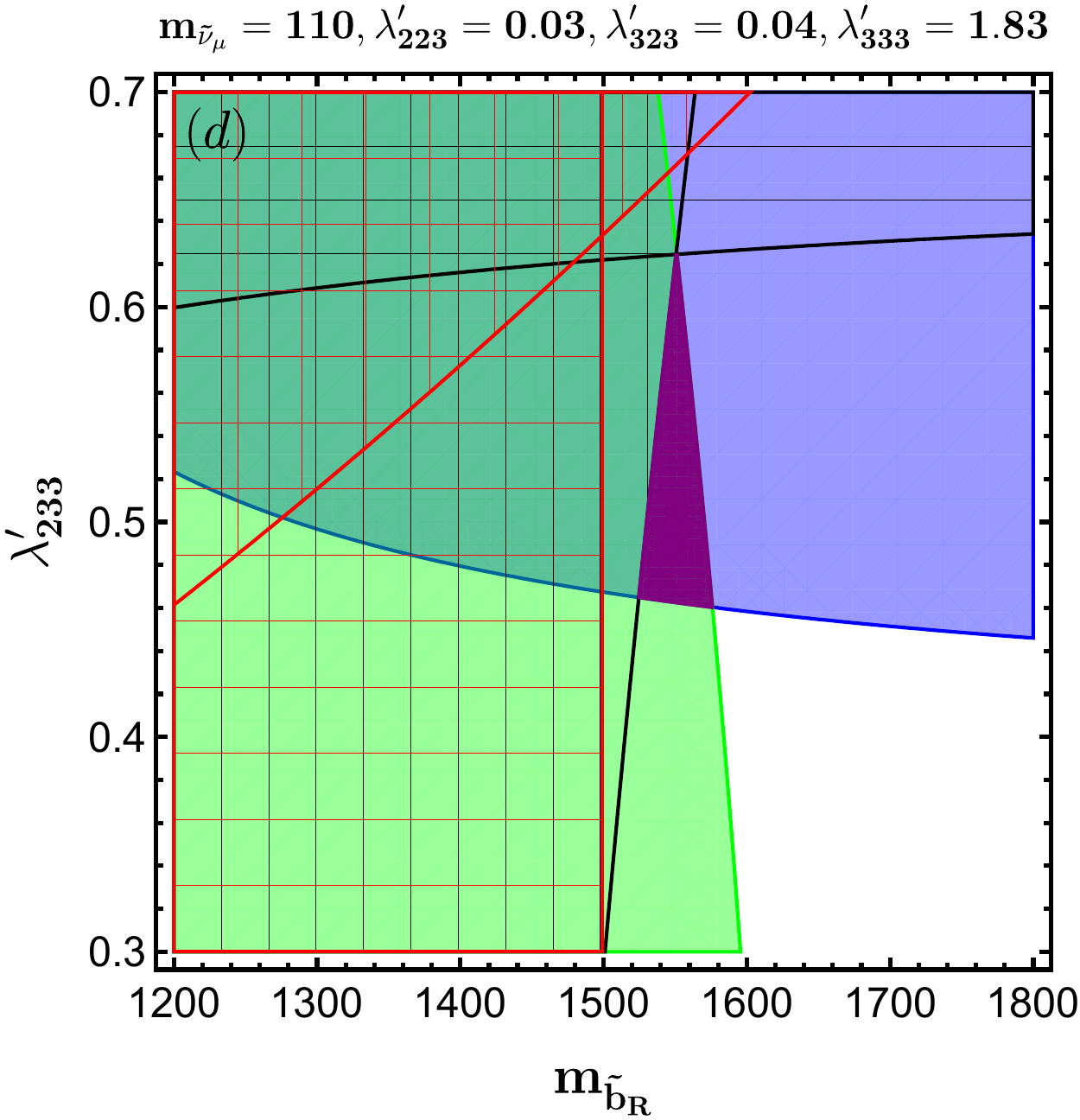}\quad
	\includegraphics[width=0.32\textwidth]{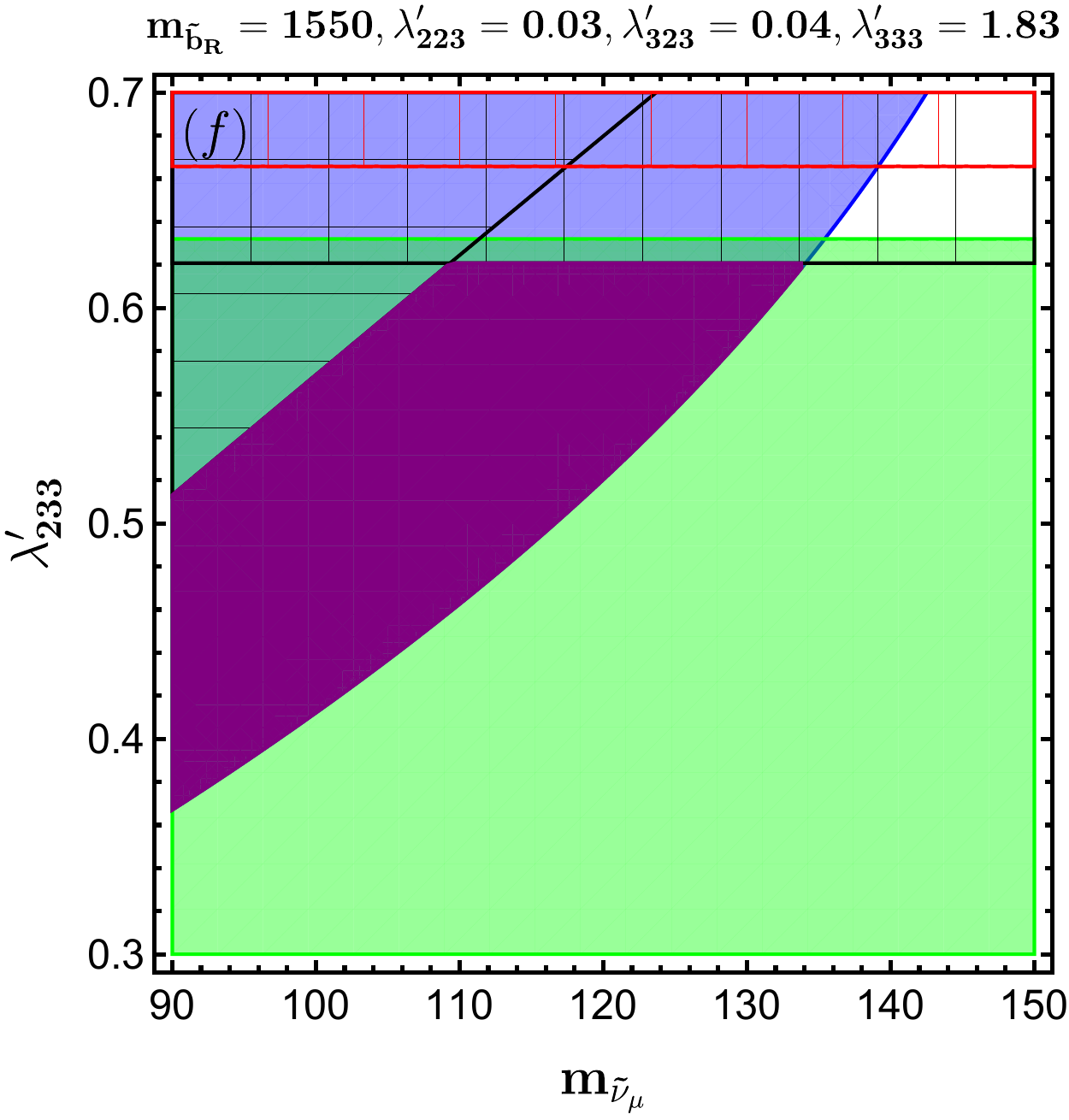}
	\caption{\label{fig:results}Numerical analysis in which $b \to s \ell^+ \ell^-$ and $R(D^{(\ast)})$ anomalies are solved and other constraints are satisfied. The masses $m_{\tilde{b}_R}$ and $m_{\tilde{\nu}_{\mu}}$ are given in units of GeV. The $2\sigma$ favored regions from the $b \to s \ell^+ \ell^-$ and $R(D^{(\ast)})$ measurements are shown in blue and green, respectively. The hatched areas filled with black-vertical, black-horizontal, red-horizontal, and red-vertical lines are excluded by $B \to K^{(\ast)} \nu \bar\nu$ decays, $B_s-\bar B_s$ mixing, $Z$ decays, and $\tau \to \mu\mu\mu$ decay, respectively. The overlaps are marked in purple.}
\end{figure*} 

\section{Numerical results and discussions}
\label{sec:result}

In this section, we discuss how to interpret both $b \to s \ell^+ \ell^-$ and $R(D^{(\ast)})$ anomalies and satisfy all these potential constraints simultaneously. The relevant model parameters in our scenario are the wino mass $m_{\tilde{W}}$, the mass of muon sneutrino $m_{\tilde{\nu}_\mu}$, the mass of right-handed sbottom $m_{\tilde{b}_R}$, as well as four nonzero couplings $\lambda'_{223}$, $\lambda'_{233}$, $\lambda'_{323}$, and $\lambda'_{333}$. We set $m_{\tilde{W}} = 250$~GeV. It can be seen from Ref.~\cite{Hu:2019ahp} that a positive product $\lambda'_{233} \lambda'^*_{223}$ is needed to explain the $b \to s \ell^+ \ell^-$ anomaly mainly through muon sneutrinos (the $C_9^{\rm V}$ part). Both $\lambda'_{323}$ and $\lambda'_{333}$ are positive to help solve $R(D^{(\ast)})$ anomaly by exchanging $\tilde{b}_R$ at tree level~\cite{Hu:2018lmk}. The combination of the choice of above couplings will naturally produce a negative $C_9^{\rm U}$, which is in line with the conclusion of the global analysis~\cite{Alguero:2019ptt}. Our numerical results are shown in Fig.~\ref{fig:results}. These results show that it is possible to explain $b \to s \ell^+ \ell^-$ and $R(D^{(\ast)})$ anomalies simultaneously at $2\sigma$ level\footnote{In order to consider the constraints from $B \to K^{(\ast)} \nu \bar\nu$, $\tau \to \mu \gamma$ and $\tau \to \mu\mu\mu$ decays at $2\sigma$ level, we get the experimental bounds (assuming the uncertainties follow the Gaussian distribution~\cite{Buttazzo:2017ixm}) $R^{\nu \bar \nu}_{23} < 3.3$, ${\cal B}(\tau \to \mu \gamma) < 5.4 \times 10^{-8}$ and ${\cal B}(\tau \to \mu\mu\mu) < 2.6\times 10^{-8}$, respectively.}. The regions of NP parameters that can solve $B$-physics anomalies are most constrained by $B \to K^{(\ast)} \nu \bar\nu$ decays, $B_s-\bar B_s$ mixing and $Z$ decays. In addition, the $\tau \to \mu\mu\mu$ decay can provide a weak constraint. We find that other related processes, such as $D^0 \to \mu^+ \mu^-$, $\tau \to \mu \rho^0$, $B \to \tau \nu$, $D_s \to \tau \nu$, $\tau \to K \nu$, and $\tau \to \mu \gamma$ decays, do not provide available constraints.

We show in Fig.~\ref{fig:results}a and Fig.~\ref{fig:results}b the allowed regions in the planes of coupling parameters $(\lambda'_{233},\, \lambda'_{333})$ and $(\lambda'_{223},\, \lambda'_{323})$ respectively when other parameters are fixed. These two subfigures show that in order to explain the $B$-physics anomalies, the coupling parameters need to satisfy the relation $\lambda'_{333}>\lambda'_{233}>\lambda'_{323}\simeq \lambda'_{223}$, and the required $\lambda'_{223}$ and $\lambda'_{323}$ are very small. Therefore, the next four subfigures in Fig.~\ref{fig:results} mainly discuss the relationships between the coupling parameters $\lambda'_{333}$ and $\lambda'_{233}$ and the masses $m_{\tilde{b}_R}$ and $m_{\tilde{\nu}_{\mu}}$. From Fig.~\ref{fig:results}a, we can see that $\lambda'_{333}$ is more constrained by $R(D^{(\ast)})$, $B \to K^{(\ast)} \nu \bar{\nu}$ and $Z$ decays, but less affected by $b \to s \ell^+\ell^-$ processes and $B_s - \bar{B}_s$ mixing. On the contrary, $\lambda'_{233}$ is greatly constrained by $b \to s \ell^+\ell^-$ processes and $B_s - \bar{B}_s$ mixing, but has little influence on $R(D^{(\ast)})$, $B \to K^{(\ast)} \nu \bar{\nu}$ and $Z$ decays. As shown in Fig.~\ref{fig:results}c, after the variable parameter $m_{\tilde{b}_R}$ is added, the constraints of $\lambda'_{333}$ from $R(D^{(\ast)})$, $B \to K^{(\ast)} \nu \bar{\nu}$ and $Z$ decays will be relaxed a lot. The parameters $\lambda'_{333}$ and $m_{\tilde{b}_R}$ are highly correlated. Because we choose a smaller mass of muon sneutrino, the $B_s - \bar{B}_s$ mixing is more sensitive to $m_{\tilde{\nu}_{\mu}}$ than to $m_{\tilde{b}_R}$, which can be seen by comparing Fig.~\ref{fig:results}d with Fig.~\ref{fig:results}f. All subfigures contain parameter spaces (marked in purple) that can resolve $b \to s \ell^+ \ell^-$ and $R(D^{(\ast)})$ anomalies, and satisfy the constraints from other related processes simultaneously.

\section{Conclusions}
\label{sec:conclusion}

The recent measurements on semileptonic decays of $B$-meson suggest the existence of NP which breaks the LFU. Among them, the observables $R_{K^{(\ast)}}$ and $P'_5$ in $b \to s \ell^+ \ell^-$ processes and the $R(D^{(\ast)})$ in $B \to D^{(\ast)} \tau \nu$ decays are more striking. They are collectively called $B$-physics anomalies. In this work, we have explored the possibility of using muon sneutrinos $\tilde{\nu}_\mu$ and right-handed sbottoms $\tilde{b}_R$ to solve these $B$-physics anomalies simultaneously in $R$-parity violating MSSM.

To explain the anomalies in $b \to s \ell^+ \ell^-$ processes, we use a two-parameter scenario, where the total Wilson coefficients of NP are divided into two parts, one is the $C^{\rm V}_9$ (Noting $C^{\rm NP}_{10,\mu} = - C^{\rm V}_9$) that only contributes the muon channel and the other is the $C^{\rm U}_9$ that contributes both the electron and the muon channels. First, we scrutinize all the one-loop contributions of the superpotential terms $\lambda'_{ijk} L_i Q_j D_k^c$ to the $b \to s \ell^+ \ell^-$ processes under the assumptions $\lambda'_{ij1} = \lambda'_{ij2} = 0$ and $\lambda'_{1j3} = 0$. We find that the contribution from the $H^\pm - \tilde{b}_R$ box diagram (Fig.~\ref{fig:box}c) is missed in the literature, this contribution is usually positive, and we find that it is numerically negligible when $\tan\beta>2$. The photonic penguin induced by exchanging sneutrino can provide important contribution due to the existence of logarithmic enhancement, which has never been addressed before. This contribution is lepton flavour universal due to the SM photon, so it is natural to contribute a nonzero $C^{\rm U}_9$. 

Global analyses show that the sizable magnitude of $C^{\rm V}_9$ is needed to explain $b \to s \ell^+ \ell^-$ anomaly. However, $C^{\rm V}_9$ in the scenario with nonzero $C^{\rm U}_9$ is smaller than the one in the scenario without $C^{\rm U}_9$. With the addition of the latest measurements from the Belle collaboration, the world averages of $R(D^{(\ast)})$ are closer to the predicted values of the SM. These changes make it possible to use $\tilde{\nu}_\mu$ and $\tilde{b}_R$ to explain $b \to s \ell^+ \ell^-$ and $R(D^{(\ast)})$ anomalies, simultaneously. We also consider the constraints of other related processes in our scenario. The strongest constraints come from $B \to K^{(\ast)} \nu \bar\nu$ decays, $B_s-\bar B_s$ mixing, and the processes of $Z$ decays. Besides, $\tau \to \mu\mu\mu$ decay can provide a few constraints. The other decays, such as $D^0 \to \mu^+ \mu^-$, $\tau \to \mu \rho^0$, $B \to \tau \nu$, $D_s \to \tau \nu$, $\tau \to K \nu$, and $\tau \to \mu \gamma$, do not provide available constraints.

\section*{Acknowledgements}

This work is supported by the National Natural Science Foundation of China under Grant Nos.~11947083 and 11775092.


\begin{thebibliography}{10}

\bibitem{Aaij:2019wad}
{\scshape LHCb} collaboration, \emph{{Search for lepton-universality violation
  in $B^+\to K^+\ell^+\ell^-$ decays}},
  \href{https://doi.org/10.1103/PhysRevLett.122.191801}{\emph{Phys. Rev. Lett.}
  {\bfseries 122} (2019) 191801}
  [\href{https://arxiv.org/abs/1903.09252}{{\ttfamily 1903.09252}}].

\bibitem{Aaij:2014ora}
{\scshape LHCb} collaboration, \emph{{Test of lepton universality using
  $B^{+}\rightarrow K^{+}\ell^{+}\ell^{-}$ decays}},
  \href{https://doi.org/10.1103/PhysRevLett.113.151601}{\emph{Phys. Rev. Lett.}
  {\bfseries 113} (2014) 151601}
  [\href{https://arxiv.org/abs/1406.6482}{{\ttfamily 1406.6482}}].

\bibitem{Bordone:2016gaq}
M.~Bordone, G.~Isidori and A.~Pattori, \emph{{On the Standard Model predictions
  for $R_K$ and $R_{K^*}$}},
  \href{https://doi.org/10.1140/epjc/s10052-016-4274-7}{\emph{Eur. Phys. J.}
  {\bfseries C76} (2016) 440}
  [\href{https://arxiv.org/abs/1605.07633}{{\ttfamily 1605.07633}}].

\bibitem{Aaij:2017vbb}
{\scshape LHCb} collaboration, \emph{{Test of lepton universality with $B^{0}
  \rightarrow K^{*0}\ell^{+}\ell^{-}$ decays}},
  \href{https://doi.org/10.1007/JHEP08(2017)055}{\emph{JHEP} {\bfseries 08}
  (2017) 055} [\href{https://arxiv.org/abs/1705.05802}{{\ttfamily
  1705.05802}}].

\bibitem{Abdesselam:2019lab}
{\scshape Belle} collaboration, \emph{{Test of lepton flavor universality in $B
  \to K \ell^{+}\ell^{-}$ decays}},
  \href{https://arxiv.org/abs/1908.01848}{{\ttfamily 1908.01848}}.

\bibitem{Abdesselam:2019wac}
{\scshape Belle} collaboration, \emph{{Test of lepton flavor universality in
  ${B\to K^\ast\ell^+\ell^-}$ decays at Belle}},
  \href{https://arxiv.org/abs/1904.02440}{{\ttfamily 1904.02440}}.

\bibitem{DescotesGenon:2012zf}
S.~Descotes-Genon, J.~Matias, M.~Ramon and J.~Virto, \emph{{Implications from
  clean observables for the binned analysis of $B \to K^*\mu^+\mu^-$ at large
  recoil}}, \href{https://doi.org/10.1007/JHEP01(2013)048}{\emph{JHEP}
  {\bfseries 01} (2013) 048} [\href{https://arxiv.org/abs/1207.2753}{{\ttfamily
  1207.2753}}].

\bibitem{Descotes-Genon:2013vna}
S.~Descotes-Genon, T.~Hurth, J.~Matias and J.~Virto, \emph{{Optimizing the
  basis of $B \to K^* ll$ observables in the full kinematic range}},
  \href{https://doi.org/10.1007/JHEP05(2013)137}{\emph{JHEP} {\bfseries 05}
  (2013) 137} [\href{https://arxiv.org/abs/1303.5794}{{\ttfamily 1303.5794}}].

\bibitem{Hu:2016gpe}
Q.-Y. Hu, X.-Q. Li and Y.-D. Yang, \emph{{$B^0\to K^{\ast 0}\mu^+\mu^-$ decay
  in the Aligned Two-Higgs-Doublet Model}},
  \href{https://doi.org/10.1140/epjc/s10052-017-4748-2}{\emph{Eur. Phys. J.}
  {\bfseries C77} (2017) 190}
  [\href{https://arxiv.org/abs/1612.08867}{{\ttfamily 1612.08867}}].

\bibitem{Aaij:2015oid}
{\scshape LHCb} collaboration, \emph{{Angular analysis of the $B^{0} \to K^{*0}
  \mu^{+} \mu^{-}$ decay using 3 fb$^{-1}$ of integrated luminosity}},
  \href{https://doi.org/10.1007/JHEP02(2016)104}{\emph{JHEP} {\bfseries 02}
  (2016) 104} [\href{https://arxiv.org/abs/1512.04442}{{\ttfamily
  1512.04442}}].

\bibitem{Aaij:2013qta}
{\scshape LHCb} collaboration, \emph{{Measurement of Form-Factor-Independent
  Observables in the Decay $B^{0} \to K^{*0} \mu^+ \mu^-$}},
  \href{https://doi.org/10.1103/PhysRevLett.111.191801}{\emph{Phys. Rev. Lett.}
  {\bfseries 111} (2013) 191801}
  [\href{https://arxiv.org/abs/1308.1707}{{\ttfamily 1308.1707}}].

\bibitem{Khachatryan:2015isa}
{\scshape CMS} collaboration, \emph{{Angular analysis of the decay $B^0 \to
  K^{*0} \mu^+ \mu^-$ from pp collisions at $\sqrt s = 8$ TeV}},
  \href{https://doi.org/10.1016/j.physletb.2015.12.020}{\emph{Phys. Lett.}
  {\bfseries B753} (2016) 424}
  [\href{https://arxiv.org/abs/1507.08126}{{\ttfamily 1507.08126}}].

\bibitem{Aaboud:2018krd}
{\scshape ATLAS} collaboration, \emph{{Angular analysis of $B^0_d \rightarrow
  K^{*}\mu^+\mu^-$ decays in $pp$ collisions at $\sqrt{s}= 8$ TeV with the
  ATLAS detector}}, \href{https://doi.org/10.1007/JHEP10(2018)047}{\emph{JHEP}
  {\bfseries 10} (2018) 047}
  [\href{https://arxiv.org/abs/1805.04000}{{\ttfamily 1805.04000}}].

\bibitem{Wehle:2016yoi}
{\scshape Belle} collaboration, \emph{{Lepton-Flavor-Dependent Angular Analysis
  of $B\to K^\ast \ell^+\ell^-$}},
  \href{https://doi.org/10.1103/PhysRevLett.118.111801}{\emph{Phys. Rev. Lett.}
  {\bfseries 118} (2017) 111801}
  [\href{https://arxiv.org/abs/1612.05014}{{\ttfamily 1612.05014}}].

\bibitem{Abdesselam:2016llu}
{\scshape Belle} collaboration, \emph{{Angular analysis of $B^0 \to
  K^\ast(892)^0 \ell^+ \ell^-$}},  
  \href{https://arxiv.org/abs/1604.04042}{{\ttfamily 1604.04042}}.

\bibitem{Aaij:2015esa}
{\scshape LHCb} collaboration, \emph{{Angular analysis and differential
  branching fraction of the decay $B^0_s\to\phi\mu^+\mu^-$}},
  \href{https://doi.org/10.1007/JHEP09(2015)179}{\emph{JHEP} {\bfseries 09}
  (2015) 179} [\href{https://arxiv.org/abs/1506.08777}{{\ttfamily
  1506.08777}}].

\bibitem{Aaij:2013aln}
{\scshape LHCb} collaboration, \emph{{Differential branching fraction and
  angular analysis of the decay $B_s^0\to\phi\mu^{+}\mu^{-}$}},
  \href{https://doi.org/10.1007/JHEP07(2013)084}{\emph{JHEP} {\bfseries 07}
  (2013) 084} [\href{https://arxiv.org/abs/1305.2168}{{\ttfamily 1305.2168}}].

\bibitem{Aebischer:2019mlg}
J.~Aebischer, W.~Altmannshofer, D.~Guadagnoli, M.~Reboud, P.~Stangl and D.~M.
  Straub, \emph{{B-decay discrepancies after Moriond 2019}},
  \href{https://arxiv.org/abs/1903.10434}{{\ttfamily 1903.10434}}.

\bibitem{Alok:2019ufo}
A.~K. Alok, A.~Dighe, S.~Gangal and D.~Kumar, \emph{{Continuing search for new
  physics in $b \to s \mu \mu$ decays: two operators at a time}},
  \href{https://doi.org/10.1007/JHEP06(2019)089}{\emph{JHEP} {\bfseries 06}
  (2019) 089} [\href{https://arxiv.org/abs/1903.09617}{{\ttfamily
  1903.09617}}].

\bibitem{Alguero:2019ptt}
M.~Algueró, B.~Capdevila, A.~Crivellin, S.~Descotes-Genon, P.~Masjuan,
  J.~Matias et~al., \emph{{Emerging patterns of New Physics with and without
  Lepton Flavour Universal contributions}},
  \href{https://doi.org/10.1140/epjc/s10052-019-7216-3}{\emph{Eur. Phys. J.}
  {\bfseries C79} (2019) 714}
  [\href{https://arxiv.org/abs/1903.09578}{{\ttfamily 1903.09578}}].

\bibitem{Ciuchini:2019usw}
M.~Ciuchini, A.~M. Coutinho, M.~Fedele, E.~Franco, A.~Paul, L.~Silvestrini
  et~al., \emph{{New Physics in $b \to s \ell^+ \ell^-$ confronts new data on
  Lepton Universality}},
  \href{https://doi.org/10.1140/epjc/s10052-019-7210-9}{\emph{Eur. Phys. J.}
  {\bfseries C79} (2019) 719}
  [\href{https://arxiv.org/abs/1903.09632}{{\ttfamily 1903.09632}}].

\bibitem{Arbey:2019duh}
A.~Arbey, T.~Hurth, F.~Mahmoudi, D.~M. Santos and S.~Neshatpour, \emph{{Update
  on the $b \to s$ anomalies}},
  \href{https://doi.org/10.1103/PhysRevD.100.015045}{\emph{Phys. Rev.}
  {\bfseries D100} (2019) 015045}
  [\href{https://arxiv.org/abs/1904.08399}{{\ttfamily 1904.08399}}].

\bibitem{Kowalska:2019ley}
K.~Kowalska, D.~Kumar and E.~M. Sessolo, \emph{{Implications for new physics in
  $b\rightarrow s \mu \mu $ transitions after recent measurements by Belle and
  LHCb}}, \href{https://doi.org/10.1140/epjc/s10052-019-7330-2}{\emph{Eur.
  Phys. J.} {\bfseries C79} (2019) 840}
  [\href{https://arxiv.org/abs/1903.10932}{{\ttfamily 1903.10932}}].

\bibitem{Capdevila:2019tsi}
B.~Capdevila, U.~Laa and G.~Valencia, \emph{{Fitting in or odd one out? Pulls
  vs residual responses in $b\to s \ell^+\ell^-$}},
  \href{https://arxiv.org/abs/1908.03338}{{\ttfamily 1908.03338}}.

\bibitem{Bhattacharya:2019dot}
S.~Bhattacharya, A.~Biswas, S.~Nandi and S.~K. Patra, \emph{{Exhaustive Model
  Selection in $b \to s \ell \ell$ Decays: Pitting Cross-Validation against
  AIC$_c$}},  \href{https://arxiv.org/abs/1908.04835}{{\ttfamily 1908.04835}}.

\bibitem{Alguero:2018nvb}
M.~Algueró, B.~Capdevila, S.~Descotes-Genon, P.~Masjuan and J.~Matias,
  \emph{{Are we overlooking lepton flavour universal new physics in $b\to
  s\ell\ell$ ?}}, \href{https://doi.org/10.1103/PhysRevD.99.075017}{\emph{Phys.
  Rev.} {\bfseries D99} (2019) 075017}
  [\href{https://arxiv.org/abs/1809.08447}{{\ttfamily 1809.08447}}].

\bibitem{Barbier:2004ez}
R.~Barbier et~al., \emph{{R-parity violating supersymmetry}},
  \href{https://doi.org/10.1016/j.physrep.2005.08.006}{\emph{Phys. Rept.}
  {\bfseries 420} (2005) 1}
  [\href{https://arxiv.org/abs/hep-ph/0406039}{{\ttfamily hep-ph/0406039}}].

\bibitem{Lees:2012xj}
{\scshape BaBar} collaboration, \emph{{Evidence for an excess of $\bar{B} \to
  D^{(*)} \tau^-\bar{\nu}_\tau$ decays}},
  \href{https://doi.org/10.1103/PhysRevLett.109.101802}{\emph{Phys. Rev. Lett.}
  {\bfseries 109} (2012) 101802}
  [\href{https://arxiv.org/abs/1205.5442}{{\ttfamily 1205.5442}}].

\bibitem{Lees:2013uzd}
{\scshape BaBar} collaboration, \emph{{Measurement of an Excess of $\bar{B} \to
  D^{(*)}\tau^- \bar{\nu}_\tau$ Decays and Implications for Charged Higgs
  Bosons}}, \href{https://doi.org/10.1103/PhysRevD.88.072012}{\emph{Phys. Rev.}
  {\bfseries D88} (2013) 072012}
  [\href{https://arxiv.org/abs/1303.0571}{{\ttfamily 1303.0571}}].

\bibitem{Huschle:2015rga}
{\scshape Belle} collaboration, \emph{{Measurement of the branching ratio of
  $\bar{B} \to D^{(\ast)} \tau^- \bar{\nu}_\tau$ relative to $\bar{B} \to
  D^{(\ast)} \ell^- \bar{\nu}_\ell$ decays with hadronic tagging at Belle}},
  \href{https://doi.org/10.1103/PhysRevD.92.072014}{\emph{Phys. Rev.}
  {\bfseries D92} (2015) 072014}
  [\href{https://arxiv.org/abs/1507.03233}{{\ttfamily 1507.03233}}].

\bibitem{Belle:2019rba}
{\scshape Belle} collaboration, \emph{{Measurement of $\mathcal{R}(D)$ and
  $\mathcal{R}(D^*)$ with a semileptonic tagging method}},
  \href{https://arxiv.org/abs/1910.05864}{{\ttfamily 1910.05864}}.

\bibitem{Hirose:2016wfn}
{\scshape Belle} collaboration, \emph{{Measurement of the $\tau$ lepton
  polarization and $R(D^*)$ in the decay $\bar{B} \to D^* \tau^-
  \bar{\nu}_\tau$}},
  \href{https://doi.org/10.1103/PhysRevLett.118.211801}{\emph{Phys. Rev. Lett.}
  {\bfseries 118} (2017) 211801}
  [\href{https://arxiv.org/abs/1612.00529}{{\ttfamily 1612.00529}}].

\bibitem{Hirose:2017dxl}
{\scshape Belle} collaboration, \emph{{Measurement of the $\tau$ lepton
  polarization and $R(D^*)$ in the decay $\bar{B} \rightarrow D^* \tau^-
  \bar{\nu}_\tau$ with one-prong hadronic $\tau$ decays at Belle}},
  \href{https://doi.org/10.1103/PhysRevD.97.012004}{\emph{Phys. Rev.}
  {\bfseries D97} (2018) 012004}
  [\href{https://arxiv.org/abs/1709.00129}{{\ttfamily 1709.00129}}].

\bibitem{Aaij:2015yra}
{\scshape LHCb} collaboration, \emph{{Measurement of the ratio of branching
  fractions $\mathcal{B}(\bar{B}^0 \to
  D^{*+}\tau^{-}\bar{\nu}_{\tau})/\mathcal{B}(\bar{B}^0 \to
  D^{*+}\mu^{-}\bar{\nu}_{\mu})$}},
  \href{https://doi.org/10.1103/PhysRevLett.115.159901,
  10.1103/PhysRevLett.115.111803}{\emph{Phys. Rev. Lett.} {\bfseries 115}
  (2015) 111803} [\href{https://arxiv.org/abs/1506.08614}{{\ttfamily
  1506.08614}}].

\bibitem{Aaij:2017uff}
{\scshape LHCb} collaboration, \emph{{Measurement of the ratio of the $B^0 \to
  D^{*-} \tau^+ \nu_{\tau}$ and $B^0 \to D^{*-} \mu^+ \nu_{\mu}$ branching
  fractions using three-prong $\tau$-lepton decays}},
  \href{https://doi.org/10.1103/PhysRevLett.120.171802}{\emph{Phys. Rev. Lett.}
  {\bfseries 120} (2018) 171802}
  [\href{https://arxiv.org/abs/1708.08856}{{\ttfamily 1708.08856}}].

\bibitem{Aaij:2017deq}
{\scshape LHCb} collaboration, \emph{{Test of Lepton Flavor Universality by the
  measurement of the $B^0 \to D^{*-} \tau^+ \nu_{\tau}$ branching fraction
  using three-prong $\tau$ decays}},
  \href{https://doi.org/10.1103/PhysRevD.97.072013}{\emph{Phys. Rev.}
  {\bfseries D97} (2018) 072013}
  [\href{https://arxiv.org/abs/1711.02505}{{\ttfamily 1711.02505}}].

\bibitem{Amhis:2019ckw}
{\scshape HFLAV} collaboration, \emph{{Averages of $b$-hadron, $c$-hadron, and
  $\tau$-lepton properties as of 2018}},
  \href{https://arxiv.org/abs/1909.12524}{{\ttfamily 1909.12524}}.

\bibitem{Amhis:2019up}
{\scshape HFLAV} collaboration, \emph{{Online update for averages of $R_D$ and
  $R_{D^{\ast}}$ for Spring 2019 at
  \url{https://hflav-eos.web.cern.ch/hflav-eos/semi/spring19/html/RDsDsstar/RDRDs.html}}},
  .

\bibitem{Bigi:2016mdz}
D.~Bigi and P.~Gambino, \emph{{Revisiting $B\to D \ell \nu$}},
  \href{https://doi.org/10.1103/PhysRevD.94.094008}{\emph{Phys. Rev.}
  {\bfseries D94} (2016) 094008}
  [\href{https://arxiv.org/abs/1606.08030}{{\ttfamily 1606.08030}}].

\bibitem{Bernlochner:2017jka}
F.~U. Bernlochner, Z.~Ligeti, M.~Papucci and D.~J. Robinson, \emph{{Combined
  analysis of semileptonic $B$ decays to $D$ and $D^*$: $R(D^{(*)})$,
  $|V_{cb}|$, and new physics}},
  \href{https://doi.org/10.1103/PhysRevD.95.115008,
  10.1103/PhysRevD.97.059902}{\emph{Phys. Rev.} {\bfseries D95} (2017) 115008}
  [\href{https://arxiv.org/abs/1703.05330}{{\ttfamily 1703.05330}}].

\bibitem{Bigi:2017jbd}
D.~Bigi, P.~Gambino and S.~Schacht, \emph{{$R(D^*)$, $|V_{cb}|$, and the Heavy
  Quark Symmetry relations between form factors}},
  \href{https://doi.org/10.1007/JHEP11(2017)061}{\emph{JHEP} {\bfseries 11}
  (2017) 061} [\href{https://arxiv.org/abs/1707.09509}{{\ttfamily
  1707.09509}}].

\bibitem{Jaiswal:2017rve}
S.~Jaiswal, S.~Nandi and S.~K. Patra, \emph{{Extraction of $|V_{cb}|$ from
  $B\to D^{(*)}\ell\nu_\ell$ and the Standard Model predictions of
  $R(D^{(*)})$}}, \href{https://doi.org/10.1007/JHEP12(2017)060}{\emph{JHEP}
  {\bfseries 12} (2017) 060}
  [\href{https://arxiv.org/abs/1707.09977}{{\ttfamily 1707.09977}}].

\bibitem{Hu:2018veh}
Q.-Y. Hu, X.-Q. Li and Y.-D. Yang, \emph{{$b\to c\tau\nu$ transitions in the
  standard model effective field theory}},
  \href{https://doi.org/10.1140/epjc/s10052-019-6766-8}{\emph{Eur. Phys. J.}
  {\bfseries C79} (2019) 264}
  [\href{https://arxiv.org/abs/1810.04939}{{\ttfamily 1810.04939}}].
  
\bibitem{Alok:2019uqc}
A.~K. Alok, D.~Kumar, S.~Kumbhakar and S.~Uma~Sankar, \emph{{Solutions to
  $R_D$-$R_{D^*}$ in light of Belle 2019 data}},
  \href{https://doi.org/10.1016/j.nuclphysb.2020.114957}{\emph{Nucl. Phys.}
  {\bfseries B953} (2020) 114957}
  [\href{https://arxiv.org/abs/1903.10486}{{\ttfamily 1903.10486}}].

\bibitem{Murgui:2019czp}
C.~Murgui, A.~Peñuelas, M.~Jung and A.~Pich, \emph{{Global fit to $b \to c
  \tau \nu$ transitions}},
  \href{https://doi.org/10.1007/JHEP09(2019)103}{\emph{JHEP} {\bfseries 09}
  (2019) 103} [\href{https://arxiv.org/abs/1904.09311}{{\ttfamily
  1904.09311}}].

\bibitem{Shi:2019gxi}
R.-X. Shi, L.-S. Geng, B.~Grinstein, S.~Jäger and J.~Martin~Camalich,
  \emph{{Revisiting the new-physics interpretation of the $b\to c\tau\nu$
  data}}, \href{https://doi.org/10.1007/JHEP12(2019)065}{\emph{JHEP} {\bfseries
  12} (2019) 065} [\href{https://arxiv.org/abs/1905.08498}{{\ttfamily
  1905.08498}}].
  
\bibitem{Cheung:2020sbq}
K.~Cheung, Z.-R. Huang, H.-D. Li, C.-D. Lü, Y.-n. Mao and R.-Y. Tang,
  \emph{{Revisit to the $b\to c\tau\nu$ transition: in and beyond the SM}},
  \href{https://arxiv.org/abs/2002.07272}{{\ttfamily 2002.07272}}.

\bibitem{Biswas:2014gga}
S.~Biswas, D.~Chowdhury, S.~Han and S.~J. Lee, \emph{{Explaining the lepton
  non-universality at the LHCb and CMS within a unified framework}},
  \href{https://doi.org/10.1007/JHEP02(2015)142}{\emph{JHEP} {\bfseries 02}
  (2015) 142} [\href{https://arxiv.org/abs/1409.0882}{{\ttfamily 1409.0882}}].

\bibitem{Das:2017kfo}
D.~Das, C.~Hati, G.~Kumar and N.~Mahajan, \emph{{Scrutinizing $R$-parity
  violating interactions in light of $R_{K^{(\ast)}}$ data}},
  \href{https://doi.org/10.1103/PhysRevD.96.095033}{\emph{Phys. Rev.}
  {\bfseries D96} (2017) 095033}
  [\href{https://arxiv.org/abs/1705.09188}{{\ttfamily 1705.09188}}].

\bibitem{Earl:2018snx}
K.~Earl and T.~Grégoire, \emph{{Contributions to $b \rightarrow s \ell \ell$
  Anomalies from $R$-Parity Violating Interactions}},
  \href{https://doi.org/10.1007/JHEP08(2018)201}{\emph{JHEP} {\bfseries 08}
  (2018) 201} [\href{https://arxiv.org/abs/1806.01343}{{\ttfamily
  1806.01343}}].

\bibitem{Darme:2018hqg}
L.~Darmé, K.~Kowalska, L.~Roszkowski and E.~M. Sessolo, \emph{{Flavor
  anomalies and dark matter in SUSY with an extra U(1)}},
  \href{https://doi.org/10.1007/JHEP10(2018)052}{\emph{JHEP} {\bfseries 10}
  (2018) 052} [\href{https://arxiv.org/abs/1806.06036}{{\ttfamily
  1806.06036}}].

\bibitem{Hu:2019ahp}
Q.-Y. Hu and L.-L. Huang, \emph{{Explaining $b\to s \ell^+ \ell^-$ data by
  sneutrinos in the $R$-parity violating MSSM}},
  \href{https://doi.org/10.1103/PhysRevD.101.035030}{\emph{Phys. Rev.}
  {\bfseries D101} (2020) 035030}
  [\href{https://arxiv.org/abs/1912.03676}{{\ttfamily 1912.03676}}].

\bibitem{Deshpande:2012rr}
N.~G. Deshpande and A.~Menon, \emph{{Hints of R-parity violation in B decays
  into $\tau \nu$}}, \href{https://doi.org/10.1007/JHEP01(2013)025}{\emph{JHEP}
  {\bfseries 01} (2013) 025} [\href{https://arxiv.org/abs/1208.4134}{{\ttfamily
  1208.4134}}].

\bibitem{Zhu:2016xdg}
J.~Zhu, H.-M. Gan, R.-M. Wang, Y.-Y. Fan, Q.~Chang and Y.-G. Xu, \emph{{Probing
  the R-parity violating supersymmetric effects in the exclusive $b\to
  c\ell^-\bar{\nu}_\ell$ decays}},
  \href{https://doi.org/10.1103/PhysRevD.93.094023}{\emph{Phys. Rev.}
  {\bfseries D93} (2016) 094023}
  [\href{https://arxiv.org/abs/1602.06491}{{\ttfamily 1602.06491}}].

\bibitem{Altmannshofer:2017poe}
W.~Altmannshofer, P.~S. Bhupal~Dev and A.~Soni, \emph{{$R_{D^{(*)}}$ anomaly: A
  possible hint for natural supersymmetry with $R$-parity violation}},
  \href{https://doi.org/10.1103/PhysRevD.96.095010}{\emph{Phys. Rev.}
  {\bfseries D96} (2017) 095010}
  [\href{https://arxiv.org/abs/1704.06659}{{\ttfamily 1704.06659}}].

\bibitem{Hu:2018lmk}
Q.-Y. Hu, X.-Q. Li, Y.~Muramatsu and Y.-D. Yang, \emph{{R-parity violating
  solutions to the $R_{D^{(\ast)}}$ anomaly and their GUT-scale unifications}},
  \href{https://doi.org/10.1103/PhysRevD.99.015008}{\emph{Phys. Rev.}
  {\bfseries D99} (2019) 015008}
  [\href{https://arxiv.org/abs/1808.01419}{{\ttfamily 1808.01419}}].

\bibitem{Wang:2019trs}
D.-Y. Wang, Y.-D. Yang and X.-B. Yuan, \emph{{$b \to c\tau\bar\nu$ decays in
  supersymmetry with $R$-parity violation}},
  \href{https://doi.org/10.1088/1674-1137/43/8/083103}{\emph{Chin. Phys.}
  {\bfseries C43} (2019) 083103}
  [\href{https://arxiv.org/abs/1905.08784}{{\ttfamily 1905.08784}}].

\bibitem{Deshpand:2016cpw}
N.~G. Deshpande and X.-G. He, \emph{{Consequences of R-parity violating
  interactions for anomalies in $\bar B\to D^{(*)} \tau \bar \nu$ and $b\to s
  \mu^+\mu^-$}},
  \href{https://doi.org/10.1140/epjc/s10052-017-4707-y}{\emph{Eur. Phys. J.}
  {\bfseries C77} (2017) 134}
  [\href{https://arxiv.org/abs/1608.04817}{{\ttfamily 1608.04817}}].

\bibitem{Trifinopoulos:2018rna}
S.~Trifinopoulos, \emph{{Revisiting R-parity violating interactions as an
  explanation of the B-physics anomalies}},
  \href{https://doi.org/10.1140/epjc/s10052-018-6280-4}{\emph{Eur. Phys. J.}
  {\bfseries C78} (2018) 803}
  [\href{https://arxiv.org/abs/1807.01638}{{\ttfamily 1807.01638}}].

\bibitem{Trifinopoulos:2019lyo}
S.~Trifinopoulos, \emph{{B -physics anomalies: The bridge between R -parity
  violating supersymmetry and flavored dark matter}},
  \href{https://doi.org/10.1103/PhysRevD.100.115022}{\emph{Phys. Rev.}
  {\bfseries D100} (2019) 115022}
  [\href{https://arxiv.org/abs/1904.12940}{{\ttfamily 1904.12940}}].

\bibitem{Bauer:2015knc}
M.~Bauer and M.~Neubert, \emph{{Minimal Leptoquark Explanation for the
  R$_{D^{(*)}}$ , R$_K$ , and $(g-2)_\mu$ Anomalies}},
  \href{https://doi.org/10.1103/PhysRevLett.116.141802}{\emph{Phys. Rev. Lett.}
  {\bfseries 116} (2016) 141802}
  [\href{https://arxiv.org/abs/1511.01900}{{\ttfamily 1511.01900}}].

\bibitem{Aaltonen:2010fv}
{\scshape CDF} collaboration, \emph{{Search for R-parity Violating Decays of
  $\tau$ Sneutrinos to $e\mu$, $\mu\tau$, and $e\tau$ Pairs in $p\bar{p}$
  Collisions at $\sqrt{s} = 1.96$ TeV}},
  \href{https://doi.org/10.1103/PhysRevLett.105.191801}{\emph{Phys. Rev. Lett.}
  {\bfseries 105} (2010) 191801}
  [\href{https://arxiv.org/abs/1004.3042}{{\ttfamily 1004.3042}}].

\bibitem{Abazov:2010km}
{\scshape D0} collaboration, \emph{{Search for sneutrino Production in $e\mu$
  Final States in 5.3 fb$^{-1}$ of $p\bar{p}$ Collisions at $\sqrt{s}$ =1.96
  TeV}}, \href{https://doi.org/10.1103/PhysRevLett.105.191802}{\emph{Phys. Rev.
  Lett.} {\bfseries 105} (2010) 191802}
  [\href{https://arxiv.org/abs/1007.4835}{{\ttfamily 1007.4835}}].

\bibitem{Aad:2015pfa}
{\scshape ATLAS} collaboration, \emph{{Search for a Heavy Neutral Particle
  Decaying to $e\mu$, $e\tau$, or $\mu\tau$ in $pp$ Collisions at $\sqrt{s}=8$
  TeV with the ATLAS Detector}},
  \href{https://doi.org/10.1103/PhysRevLett.115.031801}{\emph{Phys. Rev. Lett.}
  {\bfseries 115} (2015) 031801}
  [\href{https://arxiv.org/abs/1503.04430}{{\ttfamily 1503.04430}}].

\bibitem{Khachatryan:2016ovq}
{\scshape CMS} collaboration, \emph{{Search for lepton flavour violating decays
  of heavy resonances and quantum black holes to an e$\mu$ pair in
  proton-proton collisions at $\sqrt{s}$ = 8 TeV}},
  \href{https://doi.org/10.1140/epjc/s10052-016-4149-y}{\emph{Eur. Phys. J.}
  {\bfseries C76} (2016) 317}
  [\href{https://arxiv.org/abs/1604.05239}{{\ttfamily 1604.05239}}].

\bibitem{Altmannshofer:2014rta}
W.~Altmannshofer and D.~M. Straub, \emph{{New physics in $b\rightarrow s$
  transitions after LHC run 1}},
  \href{https://doi.org/10.1140/epjc/s10052-015-3602-7}{\emph{Eur. Phys. J.}
  {\bfseries C75} (2015) 382}
  [\href{https://arxiv.org/abs/1411.3161}{{\ttfamily 1411.3161}}].

\bibitem{Passarino:1978jh}
G.~Passarino and M.~J.~G. Veltman, \emph{{One Loop Corrections for $e^+ e^-$
  Annihilation Into $\mu^+ \mu^-$ in the Weinberg Model}},
  \href{https://doi.org/10.1016/0550-3213(79)90234-7}{\emph{Nucl. Phys.}
  {\bfseries B160} (1979) 151}.

\bibitem{deGouvea:2000cf}
A.~de~Gouvea, S.~Lola and K.~Tobe, \emph{{Lepton flavor violation in
  supersymmetric models with trilinear R-parity violation}},
  \href{https://doi.org/10.1103/PhysRevD.63.035004}{\emph{Phys. Rev.}
  {\bfseries D63} (2001) 035004}
  [\href{https://arxiv.org/abs/hep-ph/0008085}{{\ttfamily hep-ph/0008085}}].

\bibitem{Das:2016vkr}
D.~Das, C.~Hati, G.~Kumar and N.~Mahajan, \emph{{Towards a unified explanation
  of $R_{D^{(\ast)}}$, $R_{K}$ and $(g-2)_{\mu}$ anomalies in a left-right
  model with leptoquarks}},
  \href{https://doi.org/10.1103/PhysRevD.94.055034}{\emph{Phys. Rev.}
  {\bfseries D94} (2016) 055034}
  [\href{https://arxiv.org/abs/1605.06313}{{\ttfamily 1605.06313}}].
  
\bibitem{Kpatcha:2019pve}
E.~Kpatcha, I.~Lara, D.~E. López-Fogliani and C.~Muñoz, \emph{{Explaining
  muon $g-2$ data in the $\mu\nu$SSM}},
  \href{https://arxiv.org/abs/1912.04163}{{\ttfamily 1912.04163}}.

\bibitem{Buras:2014fpa}
A.~J. Buras, J.~Girrbach-Noe, C.~Niehoff and D.~M. Straub, \emph{{$ B\to
  {K}^{\left(\ast \right)}\nu \overline{\nu} $ decays in the Standard Model and
  beyond}}, \href{https://doi.org/10.1007/JHEP02(2015)184}{\emph{JHEP}
  {\bfseries 02} (2015) 184} [\href{https://arxiv.org/abs/1409.4557}{{\ttfamily
  1409.4557}}].
  
\bibitem{Grygier:2017tzo}
{\scshape Belle} collaboration, \emph{{Search for $\boldsymbol{B\to
  h\nu\bar{\nu}}$ decays with semileptonic tagging at Belle}},
  \href{https://doi.org/10.1103/PhysRevD.97.099902,
  10.1103/PhysRevD.96.091101}{\emph{Phys. Rev.} {\bfseries D96} (2017) 091101}
  [\href{https://arxiv.org/abs/1702.03224}{{\ttfamily 1702.03224}}].
  
\bibitem{Lees:2013kla}
{\scshape BaBar} collaboration, \emph{{Search for $B \to K^{(*)} \nu \overline
  \nu$ and invisible quarkonium decays}},
  \href{https://doi.org/10.1103/PhysRevD.87.112005}{\emph{Phys. Rev.}
  {\bfseries D87} (2013) 112005}
  [\href{https://arxiv.org/abs/1303.7465}{{\ttfamily 1303.7465}}].

\bibitem{Lutz:2013ftz}
{\scshape Belle} collaboration, \emph{{Search for $B \to h^{(*)} \nu \bar{\nu}$
  with the full Belle $\Upsilon(4S)$ data sample}},
  \href{https://doi.org/10.1103/PhysRevD.87.111103}{\emph{Phys. Rev.}
  {\bfseries D87} (2013) 111103}
  [\href{https://arxiv.org/abs/1303.3719}{{\ttfamily 1303.3719}}].

\bibitem{Du:2015tda}
D.~Du, A.~X. El-Khadra, S.~Gottlieb, A.~S. Kronfeld, J.~Laiho, E.~Lunghi
  et~al., \emph{{Phenomenology of semileptonic B-meson decays with form factors
  from lattice QCD}},
  \href{https://doi.org/10.1103/PhysRevD.93.034005}{\emph{Phys. Rev.}
  {\bfseries D93} (2016) 034005}
  [\href{https://arxiv.org/abs/1510.02349}{{\ttfamily 1510.02349}}].

\bibitem{Aebischer:2018iyb}
J.~Aebischer, J.~Kumar, P.~Stangl and D.~M. Straub, \emph{{A Global Likelihood
  for Precision Constraints and Flavour Anomalies}},
  \href{https://doi.org/10.1140/epjc/s10052-019-6977-z}{\emph{Eur. Phys. J.}
  {\bfseries C79} (2019) 509}
  [\href{https://arxiv.org/abs/1810.07698}{{\ttfamily 1810.07698}}].

\bibitem{Tanabashi:2018oca}
{\scshape Particle Data Group} collaboration, \emph{{Review of Particle
  Physics}}, \href{https://doi.org/10.1103/PhysRevD.98.030001}{\emph{Phys.
  Rev.} {\bfseries D98} (2018) 030001}.

\bibitem{Aoki:2019cca}
{\scshape Flavour Lattice Averaging Group} collaboration, \emph{{FLAG Review
  2019}}, \href{https://doi.org/10.1140/epjc/s10052-019-7354-7}{\emph{Eur.
  Phys. J.} {\bfseries C80} (2020) 113}
  [\href{https://arxiv.org/abs/1902.08191}{{\ttfamily 1902.08191}}].

\bibitem{Kim:1997rr}
J.~E. Kim, P.~Ko and D.-G. Lee, \emph{{More on R-parity and lepton family
  number violating couplings from muon(ium) conversion, and tau and pi0
  decays}}, \href{https://doi.org/10.1103/PhysRevD.56.100}{\emph{Phys. Rev.}
  {\bfseries D56} (1997) 100}
  [\href{https://arxiv.org/abs/hep-ph/9701381}{{\ttfamily hep-ph/9701381}}].

\bibitem{Nandi:2016wlp}
S.~Nandi, S.~K. Patra and A.~Soni, \emph{{Correlating new physics signals in $B
  \to D^{(*)} \tau \nu_{\tau}$ with $B \to \tau \nu_{\tau}$}},
  \href{https://arxiv.org/abs/1605.07191}{{\ttfamily 1605.07191}}.

\bibitem{Bona:2007vi}
{\scshape UTfit} collaboration, \emph{{Model-independent constraints on $\Delta
  F=2$ operators and the scale of new physics}},
  \href{https://doi.org/10.1088/1126-6708/2008/03/049}{\emph{JHEP} {\bfseries
  03} (2008) 049} [\href{https://arxiv.org/abs/0707.0636}{{\ttfamily
  0707.0636}}], updates are available at \url{http://utfit.org/UTfit/WebHome}.
  
\bibitem{Arnan:2019olv}
P.~Arnan, D.~Becirevic, F.~Mescia and O.~Sumensari, \emph{{Probing low energy
  scalar leptoquarks by the leptonic $W$ and $Z$ couplings}},
  \href{https://doi.org/10.1007/JHEP02(2019)109}{\emph{JHEP} {\bfseries 02}
  (2019) 109} [\href{https://arxiv.org/abs/1901.06315}{{\ttfamily
  1901.06315}}].

\bibitem{Kuno:1999jp}
Y.~Kuno and Y.~Okada, \emph{{Muon decay and physics beyond the standard
  model}}, \href{https://doi.org/10.1103/RevModPhys.73.151}{\emph{Rev. Mod.
  Phys.} {\bfseries 73} (2001) 151}
  [\href{https://arxiv.org/abs/hep-ph/9909265}{{\ttfamily hep-ph/9909265}}].

\bibitem{Farzan:2010nh}
Y.~Farzan and S.~Najjari, \emph{{Extracting the CP-violating phases of
  trilinear R-parity violating couplings from $\mu \to eee$}},
  \href{https://doi.org/10.1016/j.physletb.2010.05.005}{\emph{Phys. Lett.}
  {\bfseries B690} (2010) 48}
  [\href{https://arxiv.org/abs/1001.3207}{{\ttfamily 1001.3207}}].
  
\bibitem{Buttazzo:2017ixm}
D.~Buttazzo, A.~Greljo, G.~Isidori and D.~Marzocca, \emph{{B-physics anomalies:
  a guide to combined explanations}},
  \href{https://doi.org/10.1007/JHEP11(2017)044}{\emph{JHEP} {\bfseries 11}
  (2017) 044} [\href{https://arxiv.org/abs/1706.07808}{{\ttfamily
  1706.07808}}].

\end{thebibliography}

\end{document}